\documentclass[longauth]{aa}  
\usepackage{graphicx}
\usepackage{txfonts}
\usepackage[colorlinks=true, citecolor=blue]{hyperref}

\usepackage{multirow}
\usepackage{pbox}
\usepackage{booktabs}
\usepackage{color}
\usepackage{pdflscape}

\newcommand{\oiii}{[\ion{O}{iii}]+H$\beta$}
\newcommand{\nii}{H$\alpha$+[\ion{N}{ii}]}
\newcommand{\oline}{[\ion{O}{iii}]}
\newcommand{\ha}{H$\alpha$}
\newcommand{\jwst}{\textit{JWST}}
\newcommand{\cii}{[\ion{C}{ii}]}
\newcommand{\hii}{\ion{H}{ii}}
\newcommand{\xion}{$\xi_{\mathrm{ion,0}}$}
\DeclareUnicodeCharacter{0301}{\'{e}}

\newcommand{\rxjz}{RXCJ0600-z6-3} %

\begin{document} 

   \title{Outshining in the Spatially Resolved Analysis of a Strongly-Lensed Galaxy at $z=6.072$ with \textit{JWST} NIRCam}

    \author{C. Gim\'{e}nez-Arteaga\inst{1,2}, 
 S.~Fujimoto\inst{3}, 
 F.~Valentino\inst{4,1}, 
 G.~B.~Brammer\inst{1,2},
 C.~A.~Mason\inst{1,2},
 F.~Rizzo\inst{1,2},
 V.~Rusakov\inst{1,2},
 L.~Colina\inst{5},
 G.~Prieto-Lyon\inst{1,2},
 P.~A.~Oesch\inst{1,2,6},
 D.~Espada\inst{7,8},
 K.~E.~Heintz\inst{1,2},
 K.~K.~Knudsen\inst{9}, 
 M.~Dessauges-Zavadsky\inst{6},
 N.~Laporte\inst{10,11},
 M.~Lee\inst{1,12}, 
 G.~E.~Magdis\inst{1,2,12},
 Y.~Ono\inst{13},
 Y.~Ao\inst{14,15},
 M.~Ouchi\inst{13,16,17,18},
 K.~Kohno\inst{19,20},
 A.~M.~Koekemoer\inst{21}}

    \institute{Cosmic Dawn Center (DAWN), Denmark
    \and Niels Bohr Institute, University of Copenhagen, Jagtvej 128, DK-2200 Copenhagen N, Denmark
    \and Department of Astronomy, The University of Texas at Austin, Austin, TX, USA
    \and European Southern Observatory, Karl-Schwarzschild-Str. 2, 85748 Garching, Germany
    \and Centro de Astrobiología (CAB), CSIC-INTA, Ctra. de Ajalvir km 4, Torrejón de Ardoz, E-28850, Madrid, Spain
     \and Department of Astronomy, University of Geneva, Chemin Pegasi 51, CH-1290 Versoix, Switzerland
     \and Departamento de F\'{i}sica Te\'{o}rica y del Cosmos, Campus de Fuentenueva, Edificio Mecenas, Universidad de Granada, E-18071 Granada, Spain
   \and Instituto Carlos I de F\'{i}sica Te\'{o}rica y Computacional, Facultad de Ciencias, E-18071 Granada, Spain
    \and Department of Space, Earth and Environment, Chalmers University of Technology, SE-412 96 Gothenburg, Sweden
    \and Aix-Marseille Universit\'{e}, CNRS, CNES, LAM (Laboratoire d’Astrophysique de Marseille), UMR 7326, 13388 Marseille, France 
   \and DTU-Space, Technical University of Denmark, Elektrovej 327, DK-2800 Kgs. Lyngby, Denmark
   \and Institute for Cosmic Ray Research, The University of
  Tokyo, 5-1-5 Kashiwanoha, Kashiwa, Chiba 277-8582, Japan
  \and Purple Mountain Observatory and Key Laboratory for Radio Astronomy, Chinese Academy of Sciences, Nanjing, PR China
    \and School of Astronomy and Space Science, University of Science and Technology of China, Hefei, PR China
   \and Kavli IPMU (WPI), The University of Tokyo, 5-1-5 Kashiwanoha, Kashiwa, Chiba 277-8583, Japan
   \and National Astronomical Observatory of Japan, 2-21-1 Osawa, Mitaka, Tokyo 181-8588, Japan
  \and Department of Astronomical Science, SOKENDAI (The Graduate University for Advanced Studies), 2-21-1 Osawa, Mitaka, Tokyo, 181-8588, Japan
       \and Institute of Astronomy, Graduate School of Science, The University of Tokyo, 2-21-1, Osawa, Mitaka, Tokyo 181-0015, Japan
    \and Research Center for Early Universe, Graduate School of Science, The University of Tokyo, 7-3-1, Hongo, Bunkyo-ku, Tokyo 113-0033, Japan
    \and Space Telescope Science Institute, 3700 San Martin Dr., Baltimore, MD 21218, USA}

\authorrunning{Gim\'{e}nez-Arteaga, Fujimoto, et al.}
\titlerunning{Outshining in a Strongly-Lensed Galaxy at $z=6.072$ with \jwst\ NIRCam}


  \abstract{We present \textit{JWST}/NIRCam observations of a strongly-lensed, sub-$L^*$, multiply-imaged galaxy at $z=6.072$, with magnification factors $\mu\gtrsim20$ across the galaxy. The galaxy has rich \textit{HST}, \textit{MUSE} and \textit{ALMA} ancillary observations across a broad wavelength range. Aiming to quantify the reliability of stellar mass estimates of high redshift galaxies, we perform a spatially-resolved analysis of the physical properties at scales of $\sim$~200~pc, inferred from SED modelling of 5 \textit{JWST}/NIRCam imaging bands covering 0.16 $\mu$m $< \lambda_\mathrm{rest} <$ 0.63 $\mu$m on a pixel-by-pixel basis. We find young stars surrounded by extended older stellar populations. By comparing \nii\ and \oiii\ maps inferred from the image analysis with our additional NIRSpec IFU data, we find that the spatial distribution and strength of the line maps are in agreement with the IFU measurements. We explore different parametric star formation history forms with \textsc{Bagpipes} on the spatially-integrated photometry, finding that a double power-law star formation history retrieves the closest value to the spatially-resolved stellar mass estimate, and other SFH forms suffer from the dominant outshining emission from the youngest stars, thus underestimating the stellar mass -- up to $\sim$0.5~dex--. On the other hand, the DPL cannot match the IFU measured emission lines. Additionally, the ionizing photon production efficiency may be overestimated in a spatially-integrated approach by $\sim$0.15~dex, when compared to a spatially-resolved analysis. The agreement with the IFU measurements implies that our pixel-by-pixel results derived from the broadband images are robust, and that the mass discrepancies we find with spatially-integrated estimates are not just an effect of SED-fitting degeneracies or lack of NIRCam coverage. Additionally, this agreement points towards the pixel-by-pixel approach as a way to mitigate the general degeneracy between the flux excess from emission lines and underlying continuum, especially when lacking photometric medium-band coverage and/or IFU observations. This study stresses the importance of studying galaxies as the complex systems that they are, resolving their stellar populations when possible, or using more flexible SFH parameterisations. This can aid our understanding of the early stages of galaxy evolution by addressing the challenge of inferring robust stellar masses and ionizing photon production efficiencies of high redshift galaxies.}

   \keywords{extragalactic astronomy --
               high-redshift galaxies --
              star forming regions --
              gravitational lensing}

   \maketitle
%
\nolinenumbers
\section{Introduction}

One of the most critical problems in the study of high-redshift galaxies is the challenge of inferring robust stellar masses. Recent spatially-unresolved works using the first data obtained with \jwst\ have found surprisingly large stellar masses, which may be in conflict with the early growth of structure within the $\Lambda$CDM cosmological model \citep{Labbe23,Xiao23}. Various studies have found that varying the star formation history (SFH) parameterisation, or other assumptions such as the initial mass function (IMF), can have a significant impact in the inferred physical properties, potentially solving this conflict \citep[see e.g.,][]{Suess22,Whitler23,Tacchella23,Pacifici23,Steinhardt23,Endsley23,Wang23,Woodrum23}.

The superb spatial resolution and sensitivity of the Near-Infrared Camera (NIRCam; \citealt{2005SPIE.5904....1R,Rieke23}) onboard \textit{JWST}, allows us to extend spatially-resolved studies to high redshifts. Previous analyses have suggested that stellar mass estimates in spatially-integrated studies could be significantly underestimated, emphasising the tension with predictions from theoretical models. This has so far been addressed at lower redshifts ($z<2.5$), such as the work by \cite{Sorba18} on a statistically significant sample of $\sim1200$ galaxies, finding that resolved stellar masses can be up to five times larger than unresolved estimates. This effect has also been recently observed at high redshift ($5<z<9$), albeit on a limited sample of five galaxies in the SMACS0723 ERO field \citep{Arteaga23}.

When resolving extended galaxies and studying their stellar populations on a pixel-by-pixel basis, one can partially disentangle the problem of \textit{outshining} \citep{Sawicki98,Papovich01,Shapley01,Trager08,Graves10,Maraston10,Pforr13,Sorba15}, where young stellar populations ($<$~10~Myr) completely dominate the integrated light, hiding underlying older stellar populations ($\gtrsim$~100~Myr), thus leading to an underestimation of the total mass of the stellar population. This is particularly a problem when the coverage is limited to the UV--optical range \citep[e.g.,][]{PaulinoAfonso22}, with redder wavelengths mitigating this effect \citep[e.g.,][]{Zibetti09,Bisigello19}. Outshining has also been studied in simulations \citep[see e.g.,][]{Narayanan23}. Given the frequent degeneracy encountered between age and dust obscuration, similar stellar mass biases can be found due to dust reddening variations within a source \citep[e.g.,][]{Smail23}.

However, using only photometric observations is not enough to unequivocally determine accurately the ``true'' mass of a galaxy, as well as the many other challenges within stellar population synthesis (SPS) modelling \citep[see e.g.,][for discussions on the multiple assumptions, degeneracies and difficulties within SPS modelling]{Conroy09,Conroy10,Pforr12,Conroy13, Maraston13}. One of the advantages of \textit{JWST} is that we can also obtain spatially-resolved \textit{spectra} of galaxies at $z>6$ with the NIRSpec Integral Field Unit (IFU, \citealt{Boker22}), providing spatial maps of emission lines and overall spectral properties \citep[e.g.,][]{Wylezalek22,Rigby23,Maiolino23}. For our purposes, the IFU spectra can provide a crucial check on spatially-resolved characteristics derived from modelling the images alone. 

In this paper, we present the NIRCam observations of a strongly-magnified galaxy at $z=6.072$ \citep{Fujimoto21,Laporte21}, observed also with NIRSpec IFU. The synergy of both instruments can allow us to perform a spatially-resolved analysis of the physical properties, as well as studying the effect on the stellar mass estimates. With the IFU data, we can disentangle the excess in the reddest photometric bands, which can be caused either by very strong emission lines from \hii\ regions associated with young, recently-formed stars, or by continuum emission by older stellar populations \citep[see e.g.,][]{Labbe13,Stark13,Smit14,Smit16,Hashimoto18,Clarke21,Strait21,Marsan22}.

\begin{figure*}[t]
\centering
\includegraphics[width=\textwidth]{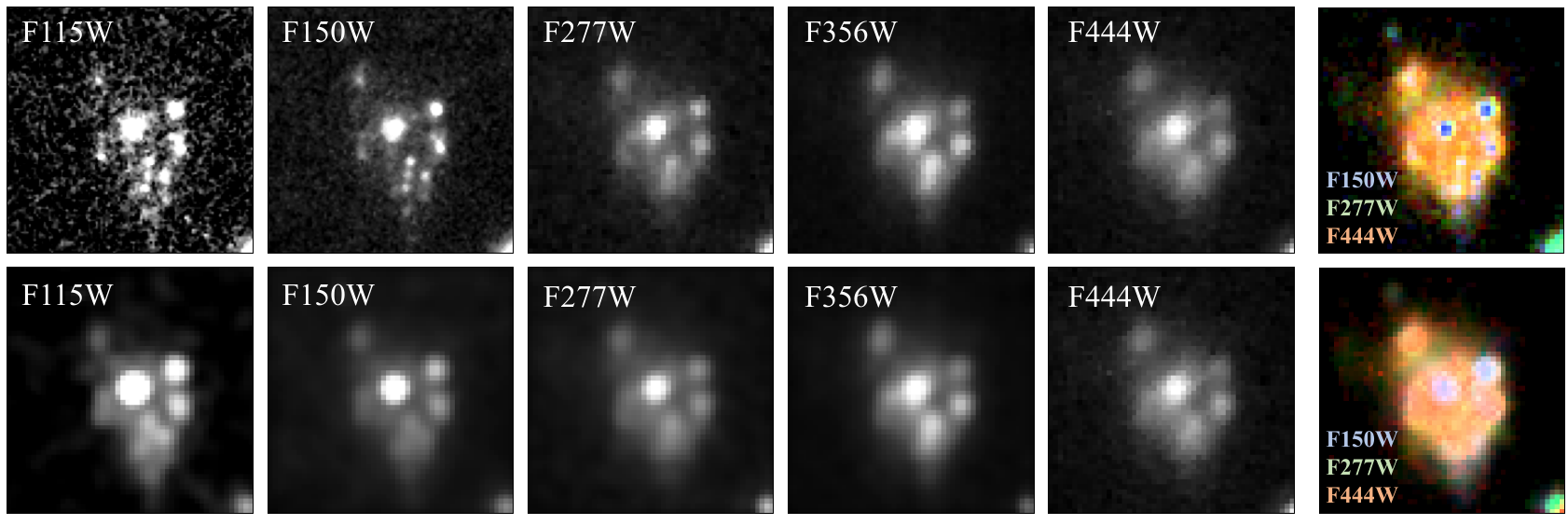}
\caption{Cutouts of the \rxjz\ image in all available NIRCam bands. The cutouts are 2~arcsec on a side and centered at $(\alpha,\delta)$~=~(06:00:09.55,$-$20:08:11.26). The top row displays the observed images in their native resolution; the bottom row shows all images convolved with a kernel to match the F444W PSF. The right RGB three-colour images are constructed from the F150W (B), F277W (G) and F444W (R) bands with the scaling following the prescription from \cite{Lupton04}. \label{fig:nircam}}
\end{figure*}

This paper is structured as follows. In Section \ref{sec:data} we present the \jwst\ NIRCam observations. Section \ref{sec:method} describes the methodology employed for the pixel-by-pixel SED modelling. In Section \ref{sec:results} we compare and discuss the results inferred from the images to ``truth'' provided by the IFU line maps. Finally, we summarise our work and present the conclusions in Section \ref{sec:conclusions}. Throughout this work, we assume a simplified $\Lambda$CDM cosmology with $H_0 = 70$ km s$^{-1}$ Mpc$^{-1}$, $\Omega_m$= 0.3 and $\Omega_{\Lambda} = 0.7$.

\section{Data} \label{sec:data}

In this work, we study a galaxy at $z=6.072$ behind the massive $z=0.43$ galaxy cluster RXCJ0600$-$2007, included in the Reionization Lensing Cluster Survey (RELICS; \citealt{Coe19}). We refer the reader to \cite{Fujimoto21} for more information on the previous multi-wavelength observations for this target, including the rest-frame UV \textit{HST} images from RELICS and far-infrared emission line and continuum measurements from the ALMA Lensing Cluster Survey (ALCS; PI: K. Kohno, Program ID: 2018.1.00035.L). In this work, we present and analyse \textit{JWST}/NIRCam photometric observations targeting this galaxy (GO-1567, PI: S. Fujimoto) obtained in January 2023, centered at (R.A., Decl.)=(06:00:05.663, $-$20:08:20.86). The exposure time varies between bands, with 2491~s for F356W and F444W, 1890~s for F115W and F277W, and a deep integration of 4982~s in F150W. The photometric data has been reduced with the \texttt{grizli} software pipeline \citep{Brammer19,grizli,grizli22}, following the procedures outlined by \cite{Valentino23} and the DAWN \textit{JWST} Archive (DJA\footnote{\url{https://dawn-cph.github.io/dja/}}). Further description of the NIRCam, NIRSpec IFU data and more recent ALMA observations will be provided by S. Fujimoto et al. (in prep.) and \cite{Valentino24}.

We correct the reduced mosaics for Milky Way extinction assuming the curve by \cite{Fitzpatrick07} and $E(B-V) = 0.0436$ \citep{MW11}. We then PSF-match all images to the F444W band using model PSFs computed with the  \texttt{WebbPSF} software \citep{Perrin12,Perrin14} and a convolution kernel to match each of the PSFs to that of F444W PSF with the \texttt{pypher} software \citep{pypher}. Finally, we resize the PSF-matched images to a common pixel scale of 40 mas~pixel$^{-1}$, which corresponds to a physical scale of $\sim$0.2 kpc~pixel$^{-1}$ at $z=6.072$.

The strongly-magnified galaxy ($\mu\sim 20-160$) has five multiple images \cite[see][and S. Fujimoto et al. in prep.]{Fujimoto21}. In this work, we focus on the image \rxjz\ ($z6.3$ in the nomenclature from \citealt{Fujimoto21}), located at (R.A., Decl.)=(06:00:09.55, $-$20:08:11.26). This image is resolved in all bands and does not show evidence of strong shears and differential magnification across the face of the source, with a magnification of $\mu = 29^{+4}_{-7}$. In contrast, the strongly-lensed arc $z6.1$/$z6.2$ is crossing the critical curve, resulting in significant variation in the very high magnification regime ($\mu\gtrsim100$) across the arc. This requires a careful magnification correction to achieve the comparison between the spatially-resolved and integrated estimates. The results obtained from the arc will be presented in Fujimoto et al. (in prep.) with the updated lens model. Moreover,  \rxjz\ traces the entire galaxy, whereas the lensing of $z6.1/z6.2$ amplifies the emission from a peripheral region of the galaxy \citep{Fujimoto21}. Finally, for the images $z6.4$ and $z6.5$ there is no IFU coverage. Figure~\ref{fig:nircam} shows the NIRCam observations on  \rxjz, for the five available broad bands F115W, F150W, F277W, F356W, and F444W, which sample roughly $\lambda_{\textrm{rest}} =$ 0.16, 0.21, 0.39, 0.50, and 0.63 $\mu$m, respectively. As can be seen in Figure~\ref{fig:nircam}, \rxjz\ has a moderate shear with the magnification variation within $\leq30\%$ across the galaxy, making it an optimal target for this study.

Throughout this work, no lensing correction is applied on any images nor derived quantities. We focus mostly on the relative differences of the inferred properties when studying them in a resolved or an integrated approach, rather than obtaining the intrinsic values for, e.g., the stellar masses or star formation rates. Therefore, all quantities are reported in terms of the magnification factor $\mu$.

\section{Methodology} \label{sec:method}

\subsection{\textsc{Bagpipes} SED Fitting} \label{sec:bagpipes}

To analyse the spatially resolved physical properties of  \rxjz, we perform spectral energy distribution (SED) fitting on a pixel-by-pixel basis. We follow the same methodology as in \cite{Arteaga23} (GA23, hereafter). We use the SED modelling code \textsc{Bagpipes} \citep{bagpipes}. We fix the redshift to the spectroscopic $z_{\textrm{\cii}}=6.072$, in order to reduce the number of free parameters in the modelling, given the limited band coverage, and break the degeneracy between redshift, age and dust. Following GA23, we use a \cite{Calzetti00} attenuation curve, and extend the nebular grid so that the ionisation parameter ($U$) can vary between $-3 < \log_{10} U < -1$, with a uniform prior in logarithmic space. The nebular emission is included with \textsc{Cloudy} \citep{cloudy}, and the SPS models are generated by \cite{BC03}. The IGM attenuation prescription assumed is from \cite{Inoue14}. We assume a \cite{Kroupa01} initial mass function (IMF), and set the lifetime of birth clouds to $t_{\rm bc}$=10~Myr. We set uniform priors for the visual extinction $A_V\in[0,3]$, and the mass of formed stars $\log_{10}(M_*/M_\odot) \in [5,11]$. Given the advantage of having IFU spectra, we use the mean metallicity measured in the cube as a prior for our SED modelling. We thus set a Gaussian prior on the metallicity centered at 0.1~$Z_\odot$, with $\sigma =$ 0.2~$Z_\odot$ (S.~Fujimoto et al. in prep.).

Throughout this work, we discuss different parameterisations of the star formation history (SFH), testing various parametric SFH forms implemented in \textsc{Bagpipes}. We are fitting a basis of four parameters, described above, as well as additional parameters for each SFHs. Table~\ref{tab:sfh_priors} in the Appendix~\ref{sec:app_sfh} provides the parametric shapes used, as well as the specific priors that we impose with each model.

\subsection{Segmentation and Pixel Selection} \label{sec:segmentation}

To perform the pixel-by-pixel SED fitting, we first need to select the pixels that will be modelled. We first use the Agglomerative Clustering package within \href{https://scikit-learn.org/stable/modules/classes.html#module-sklearn.cluster}{\texttt{sklearn.cluster}}, in order to isolate \rxjz\ and exclude nearby sources. We use the 'single' method as linkage, and apply a 1.5 distance threshold. Then, with the selected pixels associated to our desired source, we compute the S/N in all bands and apply a preliminary threshold requiring S/N~$>$~1 in all bands, on top of masking non-detections first. After that, we generate masks on each band with a new threshold that corresponds to the mean S/N on each band. Finally, we combine (i.e., sum) these masks, so that the most extended band will dominate the segmentation, and ignore pixels outside the final combined mask. The resulting S/N threshold per pixel is 1.9, 2.2, 3.5, 7.6, and 3.0 for the F115W, F150W, F277W, F356W, and F444W bands, respectively, thus ensuring that all pixels have at least a $\sim2\sigma$ detection in all bands. We obtain an image of \rxjz\ with 625 pixels that fulfil this S/N criteria, to provide a trustworthy modelling. In the results \S\ref{sec:results}, only these pixels are plotted in the various maps presented. The resulting S/N maps per band are presented in Figure~\ref{fig:app_sn} in Appendix~\ref{sec:app_sn}.

For single integrated measurements to compare to the pixel-by-pixel analysis, we sum the fluxes (and uncertainties accordingly) in each band within the same total mask and then model the integrated photometry using the same setup as for the resolved case, as explained in \S\ref{sec:bagpipes}.

\section{Results and Discussion} \label{sec:results}

In this section we present the results of the various analyses we perform with the NIRCam observations on the \rxjz\ image. We also discuss the implications of our study, from an integrated and spatially resolved perspectives. We reiterate that here we focus on the implications of inferring these properties in an unresolved versus resolved approach and do not attempt to discuss intrinsic properties affected by the lensing. A comparison with NIRSpec IFU data is presented, to test whether photometric-only estimates can reproduce spectroscopic measurements.

\subsection{Spatially Resolved Analysis} \label{sec:resolved}
 
\begin{figure*}[t]
\centering
\includegraphics[width=\textwidth]{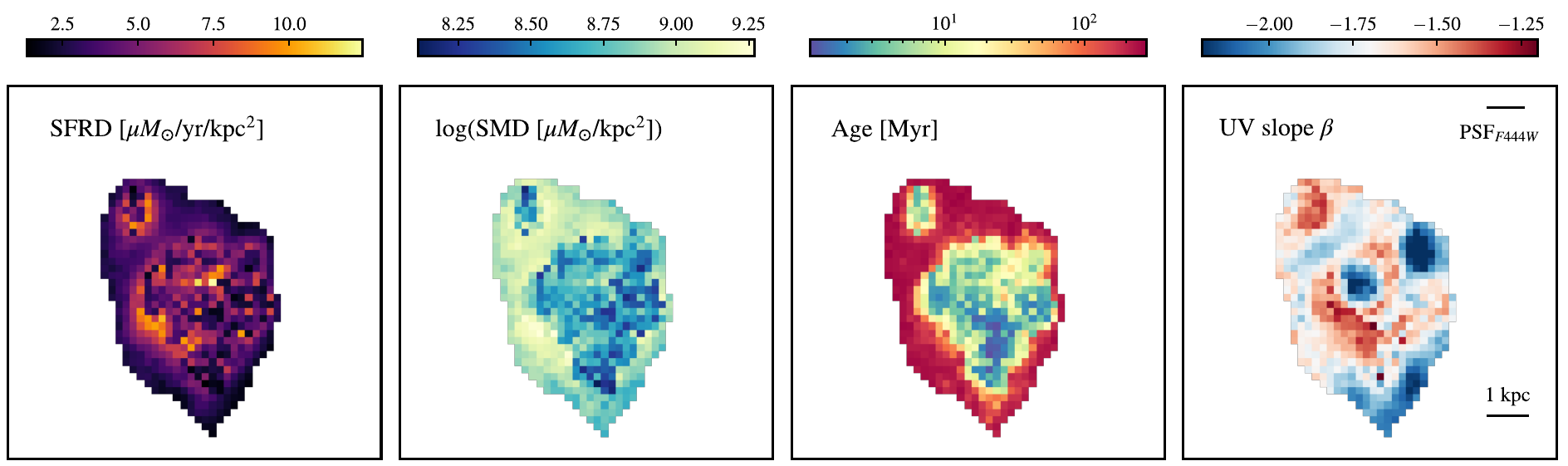}
\caption{Resulting physical properties for \rxjz\ inferred with \textsc{Bagpipes} using a constant SFH parameterisation. The maps are 2~arcsec per side and centered at $(\alpha, \delta)$=(06:00:09.55, $-$20:08:11.26). From left to right, we display the star formation rate density, the stellar mass density, the inferred mass-weighted age of the stellar population, and the UV slope ($\beta$). The size of the F444W PSF (FWHM$\sim$0\farcs16) is also indicated on the top right. \label{fig:resolved_maps}}
\end{figure*}

Following the methodology described in \S\ref{sec:method}, we perform pixel-by-pixel SED fitting on \rxjz. This allows us to build maps of the inferred physical properties. As described in \S\ref{sec:segmentation}, we only fit and display pixels that fulfil the imposed S/N criteria.

Figure \ref{fig:resolved_maps} shows the resulting maps inferred with \textsc{Bagpipes} given a constant star formation history (CSFH). This parameterisation is often used to model high-redshift galaxies \citep[e.g.,][]{Heintz23,Carnall23}. Even though a constant SFH is a simplistic parameterisation of the star formation in a galaxy, one would expect that it could be a reasonable approximation on a pixel-by-pixel basis, that would generally average over a smaller projected mix of non-coeval stellar populations. 

We obtain an equivalent scenario to the one found and discussed in GA23 for five galaxies at $5<z<9$ in the SMACS0723 field. We find centrally located clumps of young stars, surrounded by an extended region of older stellar populations. The central populations have extremely young ages of $<$10~Myr, which yields lower stellar masses, therefore we see that most mass resides in the outer shell of older stars when considering a constant star formation history. We see that the central populations are resolved, since they are larger ($\sim$$\times$3) than the size of the F444W PSF (FWHM$\sim$0\farcs16), indicated on the top right corner of the UV slope map. Additionally, the extended region is also larger than the PSF, therefore being resolved and significant given our S/N criteria. The youngest region matches an increased brightness of the clumps particularly on the F356W band (see Figure~\ref{fig:nircam}), given that this band captures the strong emission lines, as is discussed further in the following section. In the UV slope map we find two very blue clumps, which match the F115W maps, surrounded by redder slopes. The UV slope $\beta$ is inferred fitting the slope over the range $\lambda_{\rm{rest}}=1350-2800$ \AA\ on the \textsc{Bagpipes} posterior samples per pixel. In the central region, given that the age of the stellar population is relatively constant, this indicates a varying presence of dust reddening the photometry, surrounding the unobscured blue central clump. The star formation rate density, which is averaged over the last $\sim100$~Myr, is relatively constant across the galaxy, displaying the highest activity in the most massive regions.

In summary, the \rxjz\ galaxy contains centrally-concentrated clumps of very young stars ($<10$~Myr) embedded within an extended region of somewhat older stars ($>100$~Myr).

\subsection{The Integrated Field Spectrum from NIRSpec} \label{sec:ifu}

The photometry of \rxjz\ results in degenerate but distinct explanations of the rest-frame optical colors, e.g., from young stars and high equivalent-width \oiii\ and \nii\ emission lines from their associated \hii\ regions, or an underlying red continuum arising with somewhat older stars\footnote{At $z=6.072$ the groups of \oiii\ and \nii\ emission lines fall within the NIRCam F356W and F444W bandpasses, respectively.}. With \textit{JWST}/NIRSpec we can now directly observe the spatially-resolved spectra with the IFU, and test our previously presented photometric-only estimates.

\subsubsection{Strength of the Integrated Light}

From the NIRSpec IFU cube, we can directly measure the strength and spatial distribution of the emission lines \ha ($\lambda 6564\AA$), [\ion{N}{ii}] ($\lambda 6585\AA$), H$\beta$ ($\lambda 4861\AA$) and the \oline\ ($\lambda\lambda4959,5007\AA$) doublet. We refer the reader to S. Fujimoto et al. (in prep.) for all reduction, results and measurements extracted from the IFU data, both in pixel-by-pixel maps and in integrated (summed over all pixels) estimates of the equivalent width (EW) and the line fluxes.

Given the previous analysis, we can now test if our resolved estimates are consistent with the IFU spectrum. We start by summing the \textsc{Bagpipes} models for all pixels, and we calculate the EW. We consider the sum of the emission line groups \oiii\ and \nii\ as they are spectrally unresolved by the broad-band imaging filters. It is important to note that the relative strengths of the lines within each group depend on the metallicity, the ionisation parameter, amongst others, that are perhaps not fully constrained when modelling the images alone. The line fluxes and equivalent widths are listed in Table~\ref{tab:ifu}. The observed line fluxes and EWs are also measured integrating the IFU cubes within a spatial aperture radius of 0\farcs7 (S. Fujimoto et al. in prep.), and the resulting values are shown on the right column of Table~\ref{tab:ifu}.

We find that the NIRSpec IFU equivalent widths are quite large, of $\gtrsim$700~\AA\ for \oiii\ and \nii. This is consistent with what has been found in many rest-UV-selected galaxies which display strong nebular lines (EW$_{\textrm{\oiii}}\gtrsim 600$~\AA) \citep[e.g., ][]{Labbe13,Smit14, DeBarros19, Stefanon22,Rinaldi23,Caputi23}, a population that seems to rise significantly into reionization \citep{Endsley21,Whitler23}, becoming typical at $z\sim6$ \citep{Matthee23}. With the CSFH, we obtain a consistent value for EW(\nii), within the uncertainties. The NIRCam estimate for EW(\oiii) is larger than the IFU measurement, although still in agreement within $2.5\sigma$. Regarding the line fluxes, the photometric estimate is in good agreement, albeit slightly lower, with the strength of the line fluxes inferred from the NIRSpec IFU data, within the uncertainties. 

\begin{table}[t]
\small
\centering
\caption{Values for the rest-frame equivalent widths and line fluxes (in cgs units of $\times10^{-16}$ erg/s/cm$^2$) of \oiii\ and \nii\ inferred from the pixel-by-pixel modelling with \textsc{Bagpipes} on the NIRCam images, and calculated from the NIRSpec IFU measurements within a spatial aperture ratdius of 0\farcs7 (S. Fujimoto et al. in prep.).}
\begin{tabular}{l c c}
\toprule
\toprule 
 & NIRCam & NIRSpec/IFU  \\
\midrule
EW$(\textrm{\oiii})$ [\AA] & $940 \pm 20$ & $711 \pm 66$ \\
EW$(\textrm{\nii})$ [\AA] & $596 \pm 32$ & $732 \pm 125$  \\
$f_{\textrm{\oiii}}$ [cgs] & $ 2.8 \pm 0.3$ & $3.15 \pm 0.02$  \\
$f_{\textrm{\nii}}$ [cgs] & $1.1 \pm 0.2$ &  $1.44 \pm 0.02$  \\
\bottomrule
\end{tabular}
\label{tab:ifu}
\end{table}

\subsubsection{Line Emission Spatial Distribution}

Besides the integrated comparison, we can also perform a spatially-resolved analysis. While the IFU pixel scale (0\farcs1 pixel$^{-1}$) is somewhat larger than that of the NIRCam images (0\farcs04 pixel$^{-1}$), the IFU resolution is still sufficient for $\sim$300 IFU pixels across the magnified face of \rxjz.

\begin{figure}[t]
\centering
\includegraphics[width=\columnwidth]{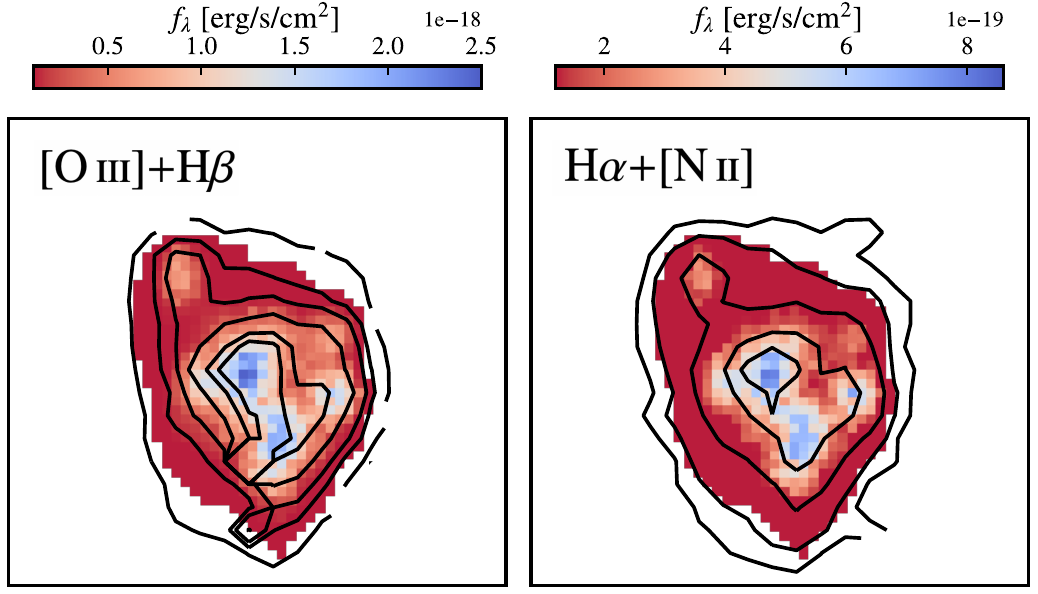}
\caption{Maps of the emission line fluxes obtained with the pixel-by-pixel SED fitting on the NIRCam images with a constant SFH. The contours correspond to the observed spectrum with the NIRSpec IFU (S. Fujimoto et al. in prep.). The maps are centered at $(\alpha, \delta)$=(06:00:09.55, $-$20:08:11.26) and 2~arcsec per side. \textbf{Left:} Map and contours for the \oiii\ emission. The IFU contour levels are 0.1, 0.2, 0.3, 0.5, 0.8, and $1\times10^{-17}$ erg s$^{-1}$ cm$^{-2}$. \textbf{Right:} Map and contours for the \nii\ emission. The contour levels correspond to 0.3, 0.5, 1, 2, and $3\times10^{-18}$ erg s$^{-1}$ cm$^{-2}$. \label{fig:ifu_maps}}
\end{figure}

In Figure~\ref{fig:ifu_maps} we show the spatially-resolved comparison between the NIRSpec IFU and NIRCam \nii\ and \oiii\ emission lines. The underlying maps correspond to the line fluxes inferred from the pixel-by-pixel SED modelling of the NIRCam data with a CSFH. The overplotted contours are of the moment 0 (intensity) measurements of the emission line groups from the IFU cubes. The NIRCam maps consistently reproduce the spatial distribution of the IFU measured lines, as well as their strength, as has also been tested from the spatially-integrated perspective. The IFU confirms the presence of strong nebular emission in the central region of \rxjz.

With these quantitative and qualitative comparisons, we conclude that the pixel-by-pixel SED modelling with \textsc{Bagpipes} using 5 photometric NIRCam bands reproduces the spatial distribution and strength of the \nii\ and \oiii\ emission lines inferred from the NIRSpec IFU observations. This gives confidence in our results, given that the IFU spectra confirms the presence of strong line emission in the central region, where then the SED software chooses younger ages of the stellar population to model the excess in photometry. The outskirts present weaker emission lines, and that yields older stellar populations. 

A spatially-resolved analysis with only broad-band photometry, has the potential to bypass the degeneracies that not having medium-bands or IFU observations introduces on single-aperture photometric fits, as well as providing a more complete picture of the internal structure and properties of galaxies. Having confidence in our spatially resolved estimates, we can now study and discuss their implications.

\subsection{The Spatial Resolution Effect on the Stellar Mass} \label{sec:mass}

Analogous to GA23, we wish to test if having spatially-resolved observations affects the inferred physical parameters of a particular galaxy, when compared to spatially-integrated studies. For this, as explained in \S\ref{sec:segmentation}, we add the pixel-by-pixel photometry in each band to create a spatially-integrated measurement with only the pixels that fulfil the above-mentioned S/N criteria, and be able to compare both scenarios one-to-one. 

Earlier studies have found that stellar masses can be underestimated with spatially-integrated fits, given how a given model parameterization of the total light may not have sufficient flexibility to fully explain complex distributions of, e.g., dust attenuation and SFH within a single galaxy. One of these studies is \cite{Sorba18} (SS18 hereafter), that used an exponentially declining SFH, finding a stellar mass discrepancy of factors up to 5. Using a constant SFH, GA23 found that the stellar mass can be underestimated $\sim$0.5--1~dex in an unresolved fit, due to outshining from the young stellar populations \citep[see also e.g.,][]{Narayanan23}. Here we test if this is also the case for \rxjz, or if we can reproduce the resolved CSFH with different SFH shapes in the integrated photometry. It is important to note that the wavelength range that is covered, particularly the lack of IR data or the coverage of the rest-frame 1--1.6~$\mu$m where the emission from older stellar populations peaks, additionally plays an important role in the inferred stellar masses and parameters \citep[see e.g.,][]{Zibetti09,Bisigello19,Song23}.

We model the integrated photometry with \textsc{Bagpipes} using the same setup as described in \S\ref{sec:bagpipes}. We test additional parametric SFH forms available within \textsc{Bagpipes} (see Appendix~\ref{sec:app_sfh}), and infer the integrated physical parameters. We choose a double-power law (DPL) SFH, which has been used to model high-redshift post-starburst galaxies \citep[see e.g.,][]{Strait23}. We also test a log-normal SFH \citep{Abramson15,Abramson16}, and an exponentially declining SFH (SS18). 

\begin{figure}[t]
\centering
\includegraphics[width=\columnwidth]{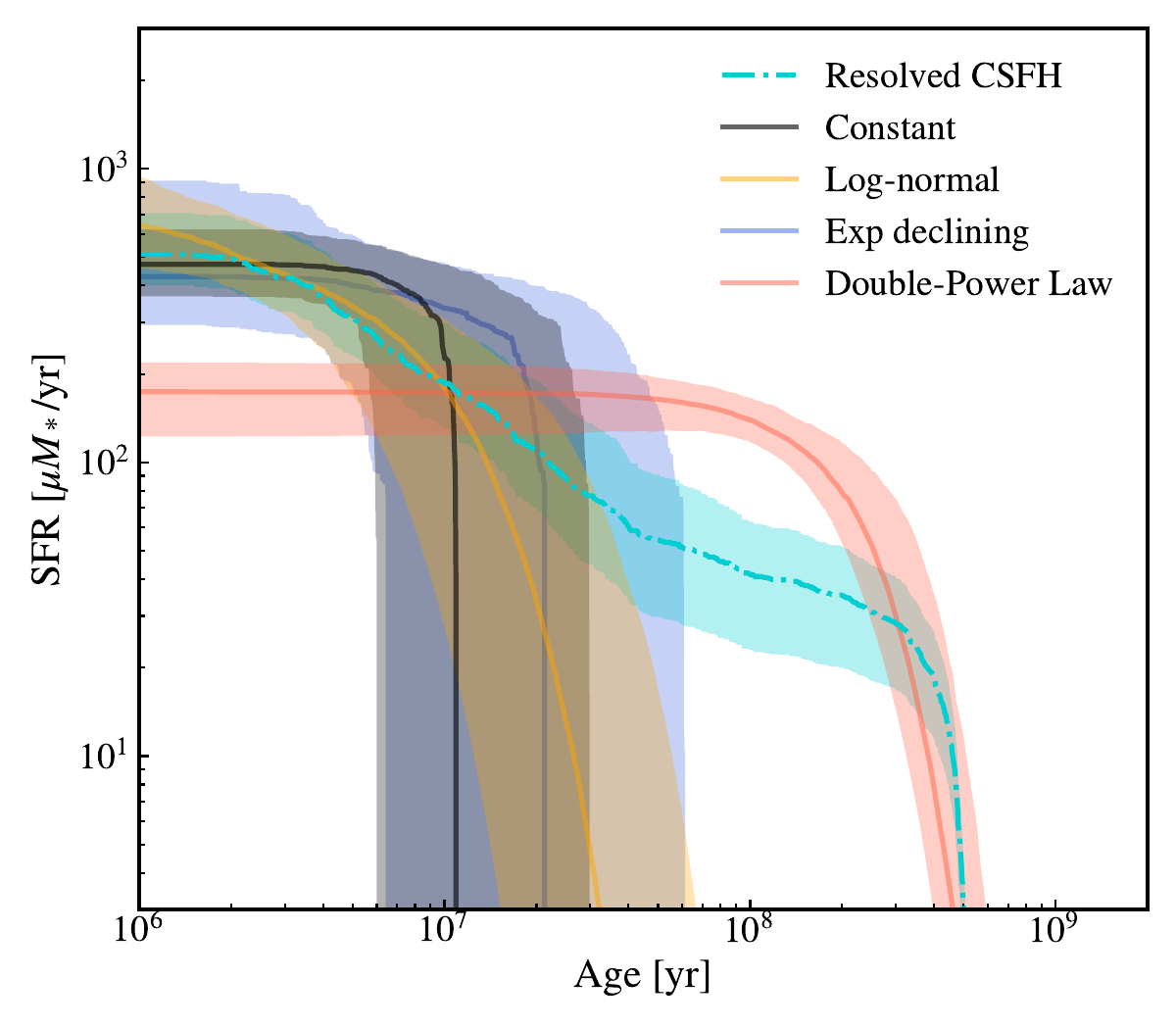}
\caption{Star formation history of the resolved constant SFH analysis (turquoise dash-dotted curve), compared with the resulting SFH of the spatially-integrated fits obtained by varying the SFH form. The shaded areas correspond to the 16–84th percentile range in each case. \label{fig:sfh_curves}}
\end{figure}

Figure \ref{fig:sfh_curves} shows the resulting SFH curves for the different models. The summed star formation histories of all individual pixels with CSFH (turquoise dashed-dotted curve and shaded region) show that the total stellar population is built up over extended timescales. On the other hand, all integrated models except the double-power law have a short, recent burst of star formation that dominates the light, but that has a lower total mass-to-light ratio. The DPL model gets closest to the summed CSFH, with older stellar populations dominating the star formation activity over long timescales, and an inferred age of $130^{+58}_{-27}$~Myr. Galaxy formation simulations predict that galaxies spend most of their time undergoing bursts of star formation, with the typical bursts having a width of $\sim100$~Myr at $z\sim6$ \citep{Ceverino18}.

The stellar masses derived with each spatially-integrated SFH form are reported in Table~\ref{tab:int_sfh}. The unresolved masses vary within log($\mu M_*/M_{\odot}$)=$9.7$ and 10.4 by changing the SFH, i.e., a difference of 0.7~dex depending on the choice of SFH form, consistent with what has been found in previous studies regarding the effect of SFH choice on the stellar masses \citep[see e.g., ][]{Whitler23,Tacchella23}.

We define the difference between the logarithmic resolved CSFH and integrated stellar mass estimates as the ``mass offset'' or $\Delta M_r$. If we add up (sum) the mass for all pixels in the resolved map presented in Figure~\ref{fig:resolved_maps}, we obtain a resolved stellar mass of log($\mu M_*/M_{\odot}$)=$10.2^{+0.2}_{-0.1}$. The mass offset between the CSFH resolved estimate and the additional SFH parameterisations on the integrated modelling are shown in Table \ref{tab:int_sfh}. The parametric shape that provides the smallest mass offset is the double-power law SFH, differing $\sim 0.2$ dex from the constant SFH resolved case (a \textit{larger} stellar mass than the resolved case), reaching no offset when considering uncertainties. The largest $\Delta M_r$ is given by all the rest of SFH integrated runs, with a significant offset of 0.5~dex, so that the resolved mass is more than 3 times larger than the unresolved one. These models, as seen also in Figure~\ref{fig:sfh_curves}, suggest very young burst of star formation, with ages of $6^{+9}_{-3}, 6^{+9}_{-4}$, and $4^{+5}_{-2}$~Myr, for a constant, log-normal, and exponentially declining SFH, respectively. None of these parameterisations allows for any significant population of stars with ages greater than $\sim$50~Myr, that are apparent in the spatially-resolved analysis, yielding mass weighted ages of $\gtrsim 100$~Myr (Figure~\ref{fig:resolved_maps}).

\begin{table}[t]
\small
\centering
\caption{Values for the stellar mass inferred in a spatially-integrated fit with \textsc{Bagpipes} using different star formation history parameterisations. The mass offset $\Delta M_r$ is calculated with respect to the resolved CSFH stellar mass of log($\mu M_*/M_{\odot}$)=$10.2^{+0.2}_{-0.1}$, thus $\Delta M_r$=log$(M_{*}^{resolved})-$log$(M_{*}^{integrated})$}.
\begin{tabular}{l c c}
\toprule
\toprule
SFH & log($\mu M_*/M_{\odot}$) & $\Delta M_r$ [dex]  \\
\midrule
Constant & $9.7^{+0.4}_{-0.2}$ & $0.5^{+0.2}_{-0.1}$ \\ 
Exponentially declining & $9.7^{+0.3}_{-0.2}$ & $0.5^{+0.2}_{-0.1}$ \\
Log-normal & $9.7^{+0.3}_{-0.2}$ & $0.5^{+0.2}_{-0.1}$ \\
Double-Power Law & $10.4^{+0.4}_{-0.2}$ & $-0.2^{+0.3}_{-0.2}$ \\
\bottomrule
\end{tabular}
\label{tab:int_sfh}
\end{table}

\begin{figure}[t]
\centering
\includegraphics[width=\columnwidth]{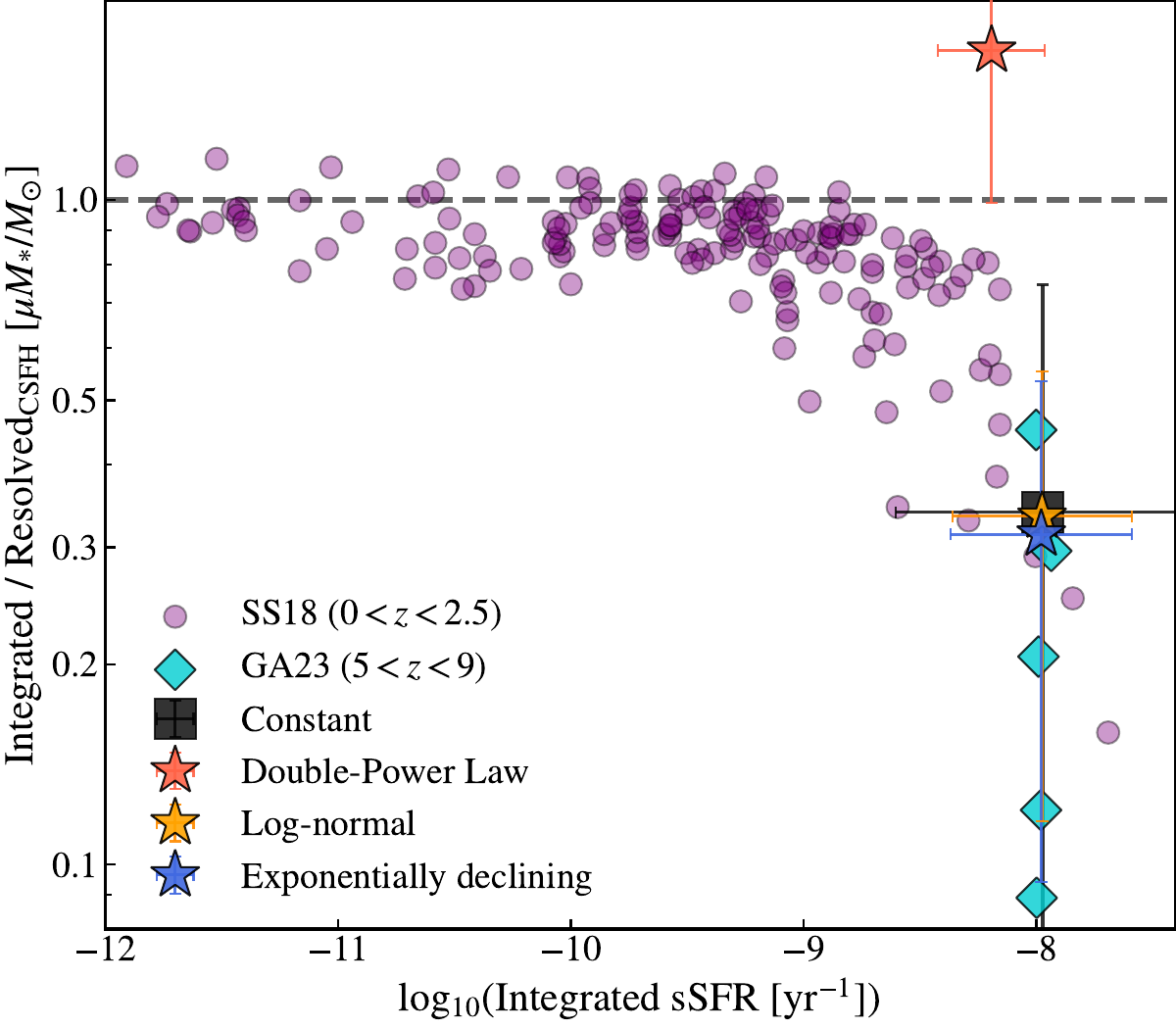}
\caption{Mass discrepancy between the integrated cases and the resolved CSFH, as a function of the unresolved specific star formation rate. The dashed line indicates no offset (or ratio=1). The purple points correspond to spectroscopically-confirmed galaxies at $0<z<2.5$ from \cite{Sorba18} (using an exponentially declining SFH). The turquoise diamonds are $5<z<9$ galaxies from \cite{Arteaga23} (using a CSFH). The square and stars correspond to this work. The black square indicates the fiducial CSFH case. The red, orange and blue stars indicate a double-power law, a log-normal, and an exponentially declining SFH parameterisations, respectively. \label{fig:sawicki}}
\end{figure}

In Figure~\ref{fig:sawicki}, we reproduce the comparison made by \citealt{Sorba18}, plotting the mass offset as a function of the integrated specific star formation rate (sSFR). Points from SS18 are for galaxies at $z_\mathrm{spec} < 2.5$ in the Hubble eXtreme Deep Field (XDF; \citealt{xdf}) with \textit{HST} images in nine bands. We add the SMACS0723 galaxies from \cite{Arteaga23} to the comparison, which have $5<z_\mathrm{spec}<9$ from \textit{JWST}/NIRSpec, and similar NIRCam images as those used in this work. Our results for \rxjz\ are shown as the black square for the CSFH, and stars for the additional SFHs parameterisations, indicated with the different colours. 

The mass discrepancies of all high redshift galaxies ($z>5$, GA23 and this work) cover the tail of the SS18 sample, which corresponds to galaxies at lower redshifts with similarly high sSFR (SS18). Our constant SFH resolved model (black square) falls in the regime of the GA23 targets, which used the same modelling prescription. Additionally, the log-normal and exponentially declining SFH agree with this discrepancy. On the other hand, the double power-law surpasses the ratio = 1 line, which means that the integrated mass is larger than the resolved one. This is not surprising given that on Figure~\ref{fig:sfh_curves} we already saw that this SFH is as extended as the resolved CSFH, with larger SF in the older stars regime. This allows for the presence of older stellar populations in the integrated fit, bypassing the outshining caused by the youngest stars that affect the other SFH parameterisations. This is also a feature of non-parametric SED-fitting codes \citep[e.g.,][]{Leja17,Leja19,Iyer20}, which also tend to infer older and more massive galaxies \citep[see e.g.,][]{Carnall19,Tacchella23}. Thus, more complex SFHs could also help disentangle the different stellar populations, as seen most recently in \cite{Jain24}.

Given the NIRCam observations alone, one might conclude that a double-power law or a SFH parameterisation that includes older stellar populations (e.g. non-parametric SFH), is a good model for targets like \rxjz, in the case that we did not have enough resolution to perform a spatially-resolved analysis. For this target, we have complementary IFU observations. From the cube, as shown in \S\ref{sec:ifu}, we find strong nebular emission in this galaxy. Given the power-law decline of the star formation rate in the DPL parameterisation, we can expect that it cannot produce extreme line emission. On top of this, if we integrate over the last $\sim10$~Myr the DPL SFH curve (Figure~\ref{fig:sfh_curves}), we can see that it will not be able to reproduce the strong line fluxes observed by the IFU such as H$\alpha$, which traces the most recent star formation. Thus, only by combining the NIRCam spatially-resolved analysis with the NIRSpec IFU measurements of the emission lines, we can see that the DPL cannot explain the strength of the lines, thus not being a good model to represent this galaxy. Moreover, the DPL presents the largest reduced chi squared value from all fits (see Figure~\ref{fig:app_fits} in Appendix~\ref{sec:app_fits}), with the rest of models providing better fits with $\chi^2_{\nu}\sim1$. Additionally, all the rest of models have $\sim0.5$~dex lower stellar mass when inferred in a spatially-integrated analysis, demonstrating once more the effect of outshining by the youngest stellar populations, as seen in previous studies \citep{Sorba18,Arteaga23,Narayanan23}, as well as the need for more complex SFHs. The implications of outshining are being studied at various redshifts \citep[e.g.,][]{Borsani20,Tang22,Topping22}. With new NIRCam data where galaxies are resolved at $z>5$, a spatially-resolved strategy can provide more adequate estimates for parameters such as the stellar mass, SFR and sSFR.

\subsection{The Ionizing Photon Production Efficiency}

Because our target is a representative galaxy ($\simeq$sub-$L^{\star}$, low-mass) within the population that is thought to be the dominant agent facilitating reionization \citep[e.g.,][]{Atek23}, it is important to study its contribution to this cosmic epoch, when reionization is thought to be almost completed \citep[$z\sim6$,][]{Fan06,Mason18}. We thus calculate the Lyman continuum (LyC) photon production efficiency ($\xi_{\mathrm{ion}}$) of \rxjz. To infer the production rate of ionizing photons that did not escape the galaxy \citep[see e.g.,][]{Prieto-Lyon23}, $\xi_\mathrm{ion,0}$, we use the equation by \cite{Bouwens16}:

\begin{equation}
    \xi_\mathrm{ion,0} \equiv \xi_\mathrm{ion} (1-f_{\mathrm{esc}}) = \frac{L_{\mathrm{H}\alpha}^{\mathrm{corr}}}{L_{\mathrm{UV}}/f_{\mathrm{esc,UV}}} \times 7.35 \times 10^{11} [\mathrm{Hz}\,\mathrm{erg}^{-1}]
    \label{eq_xion}
\end{equation}

\begin{figure}[t]
\centering
\includegraphics[width=\columnwidth]{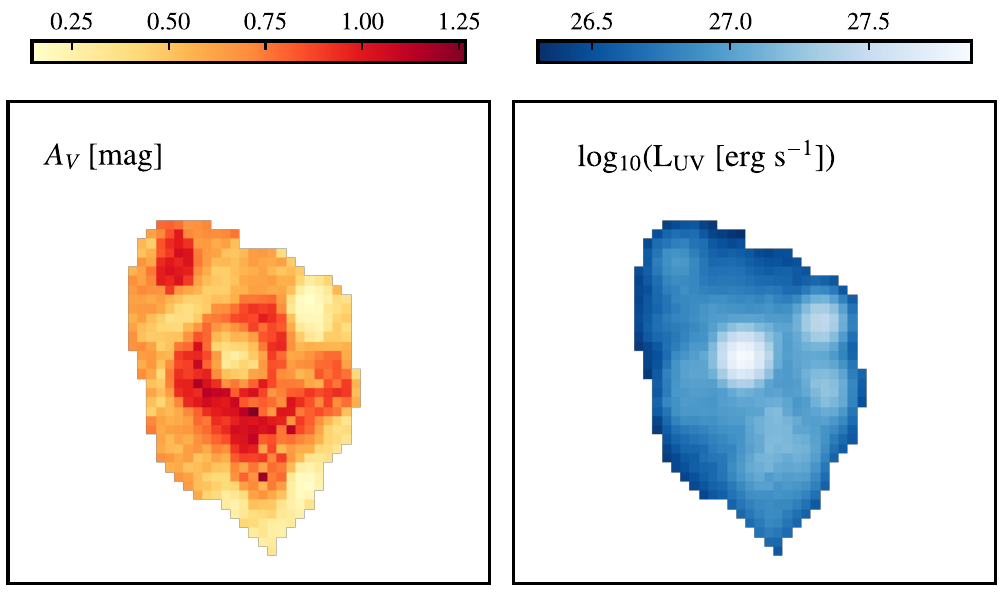}
\caption{Maps of the $A_V$ and log$_{10}(L_{\mathrm{UV}}$) inferred from the \textsc{Bagpipes} SED modelling and posterior distributions, using a CSFH.
\label{fig:av_Luv}}
\end{figure}

\noindent where $L_{\mathrm{H}\alpha}^{\mathrm{corr}}$ is the dust-corrected \ha\ luminosity, $L_{\mathrm{UV}}$ is the UV luminosity evaluated at 1500~\AA, and $1/f_{\mathrm{esc,UV}}$ accounts for the dust correction to obtain the intrinsic UV luminosity, before being obscured by dust. Using a \cite{Calzetti00} attenuation curve, we employ the relation $A_{\mathrm{UV}} = 1.99\,(\beta +2.23)$, where $\beta$ is the UV slope. From $A_{\mathrm{UV}}$ we can infer $f_{\mathrm{esc,UV}}=10^{-A_{\mathrm{UV}}/2.5}$. Finally, from the H$\alpha$ flux we infer the luminosity, and correct for dust using the $A_V$ estimated by the SED modelling \citep[see e.g. Eq.~3 in][]{Arteaga22}. It is important to note that we do not include the [N\,{\sc ii}] emission here, by integrating only over the H$\alpha$ line on each pixel, unlike in the previous analyses where we studied the combined \nii\ emission.

\begin{figure}[t]
\centering
\includegraphics[width=0.85\columnwidth]{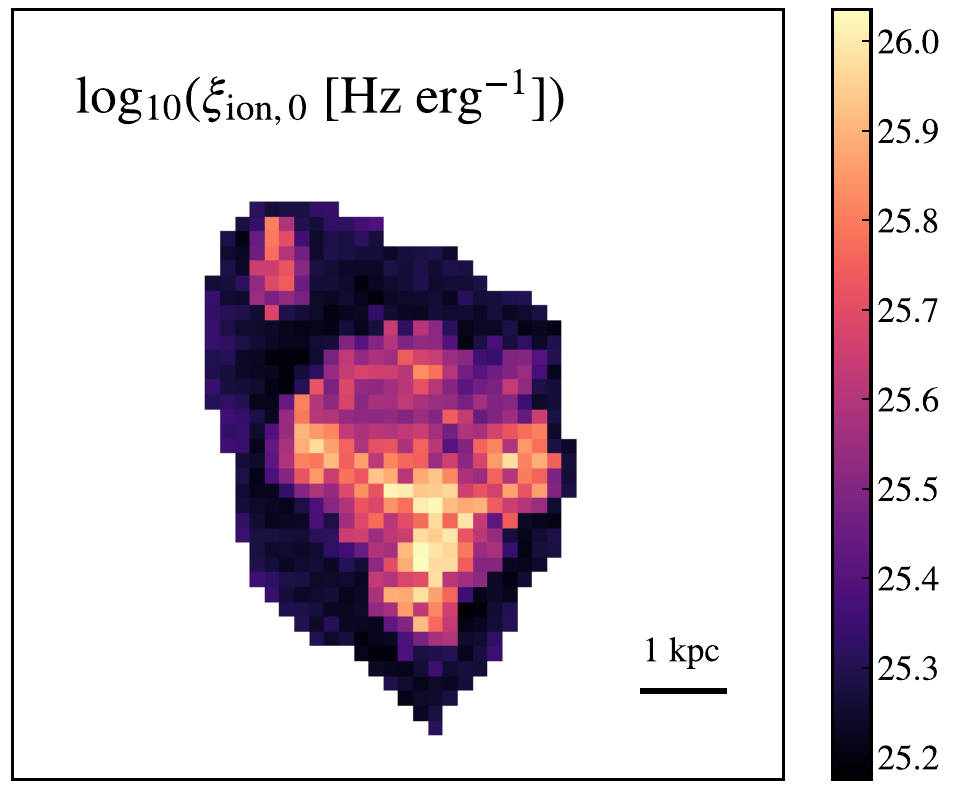}
\caption{Inferred 2D distribution of the ionizing photon production rate log$_{10}$(\xion [Hz erg$^{-1}$]), derived using Equation~\ref{eq_xion}.  
\label{fig:xion_map}}
\end{figure}

The prescription for computing \xion\ can be used on a pixel-by-pixel basis, given that we have \ha\ (Figure~\ref{fig:ifu_maps}, right panel), the UV $\beta$ slope (Figure~\ref{fig:resolved_maps}, right map), and $A_V$ and $L_{\mathrm{UV}}$ maps (Figure~\ref{fig:av_Luv}). Combining these maps according to Equation~\ref{eq_xion}, we obtain the 2D distribution of the ionizing photon production shown in Figure~\ref{fig:xion_map}. We find variations of almost an order of magnitude in \xion\ across the galaxy, from a minimum log$_{10}$(\xion [Hz/erg])=25.17, to a maximum of log$_{10}$(\xion [Hz/erg])=26.04. This could be explained by the presence of distinct stellar populations (as the spatially-resolved analysis indicates, see \S\ref{sec:resolved}), by a varying escape fraction $f_{\mathrm{esc}}$ in the different regions within the galaxy, or by a combination of both, which may be potentially correlated. A third possibility could be the presence of a hidden active galactic nuclei (AGN), which will be further addressed in Fujimoto et al. (in prep.), with the spatially-resolved optical line properties with NIRSpec IFU. As expected, given the strong correlation between age and \xion, the region where most ionizing photons are coming from corresponds to the youngest inferred ages of the stellar population (see age map in Figure~\ref{fig:resolved_maps}).

\begin{figure}[t]
\centering
\includegraphics[width=\columnwidth]{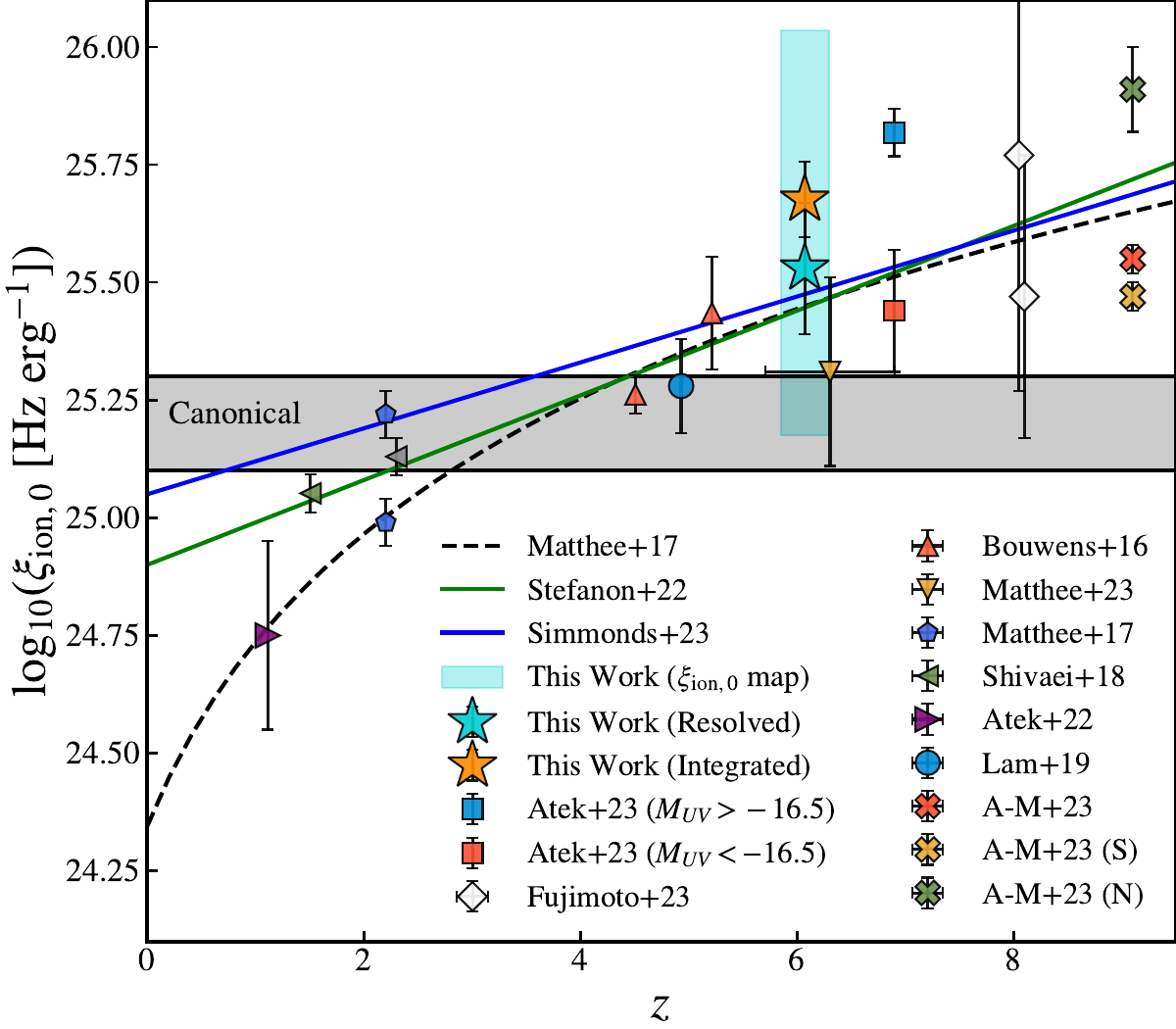}
\caption{Evolution of the ionizing photon production with redshift. The stars correspond to this work, with the orange and turquoise representing the integrated and resolved $\xi_{\mathrm{ion,0}}$ estimates, respectively. The turquoise shaded region shows the range of values of the 2D $\xi_{\mathrm{ion,0}}$ map displayed in Figure~\ref{fig:xion_map}. The grey shaded region indicates the canonical values \citep{Robertson13}. The rest of literature values are from \cite{Bouwens16,Matthee17,Shivaei18,Lam19,Atek22,Atek23,Matthee23,Fujimoto23}, \cite{AlvarezMarquez23} (A-M+23), as well as the trends from \cite{Matthee17,Stefanon22}, and \cite{Simmonds23}.
\label{fig:xi_ion}}
\end{figure}

We can additionally derive the \xion\ from our spatially-integrated photometry and estimates. Using a CSFH, to compare one-to-one with the resolved case, we obtain log$_{10}$(\xion)=$25.68\pm0.08$. Figure~\ref{fig:xi_ion} shows the evolution of the ionizing photon production rate with redshift, where we place the measurements for \rxjz, as well as additional literature values, where the mean \xion\ in the literature increases with increasing redshift. By calculating the UV-luminosity weighted mean value of the \xion\ 2D map (Figure~\ref{fig:xion_map}), we can also infer a ``resolved'' \xion\ estimate, shown in Figure~\ref{fig:xi_ion} as the turquoise star, which yields log$_{10}$(\xion)=$25.53\pm0.14$. We perform a UV-luminosity weighted mean given that \xion\ is conceptually relevant in regions where there is UV emission. Similarly to the discrepancy that we find on the stellar mass estimates, resolving the galaxy yields an offset in \xion\ of $\sim$0.15~dex, albeit the results being consistent within the uncertainties. Thus, we potentially also see the effects of outshining of 10-100~Myr stellar populations (bright in UV) by $<$10~Myr stars (brightest in H$\alpha$) in the inferred ionizing photon production efficiency, given that the youngest region dominates the inferred spatially-integrated \xion. Alternatively, if we estimate the ``resolved'' \xion\ as the mass-weighted mean (instead of UV-luminosity weighted mean), we obtain log$_{10}$(\xion)=$25.37\pm0.16$, which is more discrepant with respect to the integrated value ($\sim$0.3~dex, and not consistent within 1$\sigma$), as we would expect, given the importance of outshining in the inferred stellar masses. Our result may suggest that spatially-integrated estimates could overestimate the \xion\ measurement, at least for targets such as the one presented here with multiple young star forming clumps. Additional works with larger statistical samples are necessary to study this further. 

The galaxy \rxjz\ has one of the largest \xion\ values amongst all samples displayed in Figure~\ref{fig:xi_ion} (which are all unresolved estimates). It is only surpassed by the very faint population from \cite{Atek23} (M$_{\mathrm{UV}}>-16.5$), which reach a value of log$_{10}$(\xion)=$25.8\pm 0.05$; potentially by the galaxies from \cite{Fujimoto23}, within the uncertainties; and finally by the north clump in the galaxy MACS1149-JD1 at $z=9.1$ from \cite{AlvarezMarquez23}, which displays a remarkable log$_{10}$(\xion)=$25.91\pm0.09$. On the other hand, the lower resolved estimate is consistent with the fits derived by \cite{Stefanon22}, \cite{Matthee17} and \cite{Simmonds23}. \cite{Stefanon22} obtained the fit from a compilation of measurements up to $z\sim8$, including the ones displayed here from the works of \cite{Bouwens16}, \cite{Matthee17}, \cite{Shivaei18}, \cite{Lam19}, and \cite{Atek22}. The trend derived by \cite{Simmonds23} is performed on a sample of 677 at $z\sim4-9$ from the JADES survey \citep{Eisenstein23} observed with NIRCam imaging. The values from \cite{Fujimoto23} are from CEERS NIRCam-selected $z\gtrsim8$ galaxy candidates, spectroscopically confirmed with NIRSpec. Additionally, albeit not displayed in the figure, both our \xion\ estimates are consistent within the uncertainties with the sample from \cite{Prieto-Lyon23}, particularly with their Lyman-$\alpha$-emitting galaxies, with a median value of log$_{10}$(\xion)=$25.39\pm 0.64$, given that the large scatter that we find across a single galaxy, is also observed within the galaxy population they study at $z\sim3-7$, likely reflecting the age diversity in both cases.

The turquoise shaded rectangle in Figure~\ref{fig:xi_ion} shows the range of values that are found in the 2D map shown in Figure~\ref{fig:xion_map}. We see that some pixels within the galaxy surpass all other literature values, and others go as low as the canonical value \citep{Robertson13}, hinting once more at the complex internal structure of this galaxy, displaying a very broad range of \xion\ values. The large differences in \xion\ may indicate that some of the regions are undergoing recent bursts of star formation (traced by stronger H$\alpha$ emission over the last few Myr), whereas others are forming stars over larger timescales ($>$100~Myr, with stronger UV emission instead), which is also found on our spatially-resolved SED fitting analysis. Differences of $\sim0.4$~dex in the \xion\ have been found between the two clumps of MACS1149-JD1 (A-M+23 in Figure~\ref{fig:xi_ion}, \citealt{AlvarezMarquez23}). The spatially-integrated estimate (orange star symbol) falls at an intermediate-to-high value, given that, albeit being dominated by the youngest stars, we also find a bright central UV clump (Figure~\ref{fig:av_Luv}, right map), which then yields a more moderate \xion\ value. As we would expect, \xion\ strongly correlates with the age of the stellar population, and the pixel-by-pixel analysis yields a broad range of stellar ages and \xion\ values. The youngest stars are responsible for most of the ionizing photon production, dominating then the integrated value. The oldest stars do not contribute as significantly to \xion, and are outshined in the spatially-integrated analysis. Therefore, a spatially-resolved analysis yields a lower value of \xion (turquoise star). The two values derived for \xion\ could imply different scenarios in terms of the contribution made by this galaxy to the end of reionisation -- from being one of the largest contributors found so far, to being consistent with the derived trends from e.g. \cite{Stefanon22}.

\subsection{Caveats} \label{sec:caveats}

Despite seemingly providing the best agreement in the SFH shape with respect to the resolved CSFH, and besides not being able to reproduce the observed IFU line strengths, the double-power law SFH parameterisation has caveats to be considered. With the priors that we impose and specify in Table~\ref{tab:sfh_priors}, the \textsc{Bagpipes} modelling has seven free parameters in total. As explained throughout, we only have five bands available with NIRCam, which may not be enough to constrain all free parameters without yielding extra degeneracies between them (see Appendix~\ref{sec:app_fits}). Similarly, this caveat applies to the log-normal and exponentially declining SFHs, which have six free parameters in the fits. Therefore, despite being one of the most basic forms, the constant SFH has five free parameters, which equals the amount of data points we have to constrain them. Albeit its simplicity, it should still be a reasonable parameterisation on a pixel-by-pixel basis, as discussed before. Given the limited amount of data points, we did not explore the possibility of implementing non-parametric SFHs on a pixel-by-pixel basis, but it is worth emphasizing the flexibility that these SFH forms provide to describe with much more complexity the star formation activity. Upcoming software combining non-parametric SFHs in a spatially-resolved approach will provide a much more detailed, and potentially more accurate, view of galactic properties.

It is worth noting that with the CSFH, most mass shows as residing in the outer extended region of older stars. Given the strong lines in the center, only young stars are \textit{seen}. In reality, this region can also be a mixture of young ($<10$~Myr) and older stellar populations ($\gtrsim$100~Myr), which would increase the total mass of the galaxy. Therefore, the ``true'' stellar mass of the galaxy is most likely \textit{larger} than what a spatially-resolved analysis can recover, making the discrepancy with respect to single-aperture photometric estimates even larger. Assuming a similar SMD in all regions, equivalent to the mean SMD per pixel in non-outshined regions (defined as mass-weighted age older than 10~Myr), the galaxy would have 0.1~dex larger stellar mass.

Finally, a change of the initial mass function can also affect the inferred stellar masses \citep{Rusakov23,Steinhardt23}, and further studies factoring in the IMF assumption, the various SFH parameterisations, and the spatial resolution effect are necessary. Moreover, as stated before, we do not correct for the magnification throughout this work. It is worth noting that the magnification factors can be different for H$\alpha$ and UV, due to differences in their intrinsic surface brightnesses, which could affect the \xion\ calculation.

\section{Summary and Conclusions} \label{sec:conclusions}

We have performed pixel-by-pixel SED modelling on the NIRCam images for \rxjz, one of the multiple images of a highly-lensed galaxy at $z=6.072$ behind the galaxy cluster RXCJ0600-2007. We have built maps of the physical properties such as the stellar mass, the star formation rate, and the age of the stellar population. We have tested various parameterisations of the star formation history, and studied the effects on the retrieved stellar masses. We have tested our photometric-only estimates with measurements from NIRSpec IFU data, aiding us in verifying the conclusions from our methodology. We have studied the potential contribution to the end of reionization of this interesting galaxy. The main results of our work are as follows:
\begin{itemize}
    \item Analogous to the findings of \cite{Arteaga23} on different objects at $5<z<9$ from the SMACS0723 field, the galaxy displays centrally-located clumps of young stellar populations embedded within extended regions of older stars.
   \item The line fluxes inferred from NIRCam photometry with a CSFH can reproduce the spatial distribution and strength of the NIRSpec IFU emission line maps. 
  \item From a spatially-integrated perspective, with photometry-only modelling we retrieve consistent values of the EW and line fluxes of \oiii\ and \nii, when compared to the IFU spectra estimates.
  \item Consistent to recent findings in observations and simulations \citep[e.g.,][]{Sorba18,Narayanan23}, outshining, where young stars dominate the integrated light hiding underlying older stellar populations, affects our stellar mass values when comparing resolved versus unresolved estimates.
 \item A double-power law SFH displays star formation over extended timescales, matching the resolved CSFH curve, but cannot reproduce the strength of the IFU emission lines. Other parametric SFH shapes yield $\sim0.5$ dex lower stellar masses than the CSFH resolved estimate. No parametric model can simultaneously match the resolved stellar mass estiamte and the IFU emission line strengths.
  \item The ionizing photon production efficiency estimate may additionally be affected by outshining. Resolving the galaxy yields $\sim$0.15~dex lower \xion, which could hint at spatially-integrated studies overestimating the contribution to reionization of targets similar to the one studied here.
\end{itemize}

The problem of outshining is being studied in observations and simulations. Broadly covered in wavelength and larger statistical samples are needed to investigate this problem across cosmic time, and constrain better the systematics that it introduces in our physical estimates. The work presented here gives confidence to photometric-only spatially-resolved analyses, but stresses the importance of combining these with NIRSpec IFU or spectroscopic data, in order to break photometric degeneracies and constrain the physical properties with more accuracy.

This study puts emphasis on the importance of studying galaxies as complex systems. Given the unprecedented resolution of the instruments on board \jwst, resolving the stellar populations of galaxies, when possible, can give us different views on their internal structure and properties, when compared to the well established aperture photometry approach. By addressing the challenge of inferring robust stellar masses of high redshift galaxies, resolved studies can aid our understanding of the first stages of mass assembly and galaxy evolution.

\begin{acknowledgements}
We thank the referee for insightful comments that helped improve this article. We thank Sune Toft and Ian Smail for valuable comments and discussions. The Cosmic Dawn Center (DAWN) is funded by the Danish National Research Foundation under grant DNRF140. S.F. acknowledges the support from NASA through the NASA Hubble Fellowship grant HST-HF2-51505.001-A awarded by the Space Telescope Science Institute, which is operated by the Association of Universities for Research in Astronomy, Incorporated, under NASA contract NAS5-26555. C.A.M. acknowledges support by the VILLUM FONDEN under grant 37459. L.C. acknowledges support by grant PIB2021-127718NB-10  from the Spanish Ministry of Science and Innovation/State Agency of Research MCIN/AEI/10.13039/501100011033. P.O. is supported by the Swiss National Science Foundation through project grant 200020\_207349. This work received funding from the Swiss State Secretariat for Education, Research and Innovation (SERI). D.E. acknowledges support from a Beatriz Galindo senior fellowship (BG20/00224) from the Spanish Ministry of Science and Innovation, projects PID2020-114414GB-100 and PID2020-113689GB-I00 financed by MCIN/AEI/10.13039/501100011033, project P20-00334  financed by the Junta de Andaluc\'{i}a, and  project A-FQM-510-UGR20 of the FEDER/Junta de Andaluc\'{i}a-Consejer\'{i}a de Transformaci\'{o}n Econ\'{o}mica, Industria, Conocimiento y Universidades. K.K. acknowledges support from the Knut and Alice Wallenberg Foundation. G.E.M. acknowledges the Villum Fonden research grants 13160 and 37440. Y.A. acknowledges support from the National Natural Science Foundation of China (NSFC grants 12173089, 11933011). K.K. acknowledges the support by JSPS KAKENHI Grant numbers JP17H06130, JP22H04939, and 23K20035. This work is based on observations made with the NASA/ESA/CSA \textit{James Webb Space Telescope}. The data were obtained from the Mikulski Archive for Space Telescopes at the Space Telescope Science Institute, which is operated by the Association of Universities for Research in Astronomy, Inc., under NASA contract NAS 5-03127 for \textit{JWST}. These observations are associated with program ID. 1567, as part of a Cycle 1 GO Proposal (PI: S. Fujimoto, \citealt{2021jwst.prop.1567F}). Cloud-based data processing and file storage for this work is provided by the AWS Cloud Credits for Research program. 

Facilities: \textit{JWST} (NIRCam), \textit{JWST} (NIRSpec) \\
Software: Astropy \citep{2013A&A...558A..33A,2022ApJ...935..167A}, Matplotlib \citep{Hunter:2007}, NumPy \citep{numpy}, SciPy \citep{scipy}, \texttt{grizli} \citep{Brammer19,grizli,grizli22}

\end{acknowledgements}

\bibliographystyle{aa} 

\begin{thebibliography}{108}
\expandafter\ifx\csname natexlab\endcsname\relax\def\natexlab#1{#1}\fi

\bibitem[{{Abramson} {et~al.}(2015){Abramson}, {Gladders}, {Dressler}, {Oemler}, {Poggianti}, \& {Vulcani}}]{Abramson15}
{Abramson}, L.~E., {Gladders}, M.~D., {Dressler}, A., {et~al.} 2015, \apjl, 801, L12

\bibitem[{{Abramson} {et~al.}(2016){Abramson}, {Gladders}, {Dressler}, {Oemler}, {Poggianti}, \& {Vulcani}}]{Abramson16}
{Abramson}, L.~E., {Gladders}, M.~D., {Dressler}, A., {et~al.} 2016, \apj, 832, 7

\bibitem[{{{\'A}lvarez-M{\'a}rquez} {et~al.}(2023){{\'A}lvarez-M{\'a}rquez}, {Colina}, {Crespo G{\'o}mez}, {Rinaldi}, {Melinder}, {{\"O}stlin}, {Annunziatella}, {Labiano}, {Bik}, {Bosman}, {Greve}, {Wright}, {Alonso-Herrero}, {Boogaard}, {Azollini}, {Caputi}, {Costantin}, {Eckart}, {Garc{\'I}a-Mar{\'I}n}, {Gillman}, {Hjorth}, {Iani}, {Ilbert}, {Jermann}, {Langeroodi}, {Meyer}, {Peissker}, {P{\'e}rez-Gonz{\'a}lez}, {Pye}, {Tikkanen}, {Topinka}, {van der Werf}, {Walter}, {Henning}, \& {Ray}}]{AlvarezMarquez23}
{{\'A}lvarez-M{\'a}rquez}, J., {Colina}, L., {Crespo G{\'o}mez}, A., {et~al.} 2023, arXiv e-prints, arXiv:2309.06319

\bibitem[{{Astropy Collaboration} {et~al.}(2022){Astropy Collaboration}, {Price-Whelan}, {Lim}, {Earl}, {Starkman}, {Bradley}, {Shupe}, {Patil}, {Corrales}, {Brasseur}, {N{\"o}the}, {Donath}, {Tollerud}, {Morris}, {Ginsburg}, {Vaher}, {Weaver}, {Tocknell}, {Jamieson}, {van Kerkwijk}, {Robitaille}, {Merry}, {Bachetti}, {G{\"u}nther}, {Aldcroft}, {Alvarado-Montes}, {Archibald}, {B{\'o}di}, {Bapat}, {Barentsen}, {Baz{\'a}n}, {Biswas}, {Boquien}, {Burke}, {Cara}, {Cara}, {Conroy}, {Conseil}, {Craig}, {Cross}, {Cruz}, {D'Eugenio}, {Dencheva}, {Devillepoix}, {Dietrich}, {Eigenbrot}, {Erben}, {Ferreira}, {Foreman-Mackey}, {Fox}, {Freij}, {Garg}, {Geda}, {Glattly}, {Gondhalekar}, {Gordon}, {Grant}, {Greenfield}, {Groener}, {Guest}, {Gurovich}, {Handberg}, {Hart}, {Hatfield-Dodds}, {Homeier}, {Hosseinzadeh}, {Jenness}, {Jones}, {Joseph}, {Kalmbach}, {Karamehmetoglu}, {Ka{\l}uszy{\'n}ski}, {Kelley}, {Kern}, {Kerzendorf}, {Koch}, {Kulumani}, {Lee}, {Ly}, {Ma}, {MacBride}, {Maljaars}, {Muna}, {Murphy}, {Norman},
  {O'Steen}, {Oman}, {Pacifici}, {Pascual}, {Pascual-Granado}, {Patil}, {Perren}, {Pickering}, {Rastogi}, {Roulston}, {Ryan}, {Rykoff}, {Sabater}, {Sakurikar}, {Salgado}, {Sanghi}, {Saunders}, {Savchenko}, {Schwardt}, {Seifert-Eckert}, {Shih}, {Jain}, {Shukla}, {Sick}, {Simpson}, {Singanamalla}, {Singer}, {Singhal}, {Sinha}, {Sip{\H{o}}cz}, {Spitler}, {Stansby}, {Streicher}, {{\v{S}}umak}, {Swinbank}, {Taranu}, {Tewary}, {Tremblay}, {Val-Borro}, {Van Kooten}, {Vasovi{\'c}}, {Verma}, {de Miranda Cardoso}, {Williams}, {Wilson}, {Winkel}, {Wood-Vasey}, {Xue}, {Yoachim}, {Zhang}, {Zonca}, \& {Astropy Project Contributors}}]{2022ApJ...935..167A}
{Astropy Collaboration}, {Price-Whelan}, A.~M., {Lim}, P.~L., {et~al.} 2022, \apj, 935, 167

\bibitem[{{Astropy Collaboration} {et~al.}(2013){Astropy Collaboration}, {Robitaille}, {Tollerud}, {Greenfield}, {Droettboom}, {Bray}, {Aldcroft}, {Davis}, {Ginsburg}, {Price-Whelan}, {Kerzendorf}, {Conley}, {Crighton}, {Barbary}, {Muna}, {Ferguson}, {Grollier}, {Parikh}, {Nair}, {Unther}, {Deil}, {Woillez}, {Conseil}, {Kramer}, {Turner}, {Singer}, {Fox}, {Weaver}, {Zabalza}, {Edwards}, {Azalee Bostroem}, {Burke}, {Casey}, {Crawford}, {Dencheva}, {Ely}, {Jenness}, {Labrie}, {Lim}, {Pierfederici}, {Pontzen}, {Ptak}, {Refsdal}, {Servillat}, \& {Streicher}}]{2013A&A...558A..33A}
{Astropy Collaboration}, {Robitaille}, T.~P., {Tollerud}, E.~J., {et~al.} 2013, \aap, 558, A33

\bibitem[{{Atek} {et~al.}(2022){Atek}, {Furtak}, {Oesch}, {van Dokkum}, {Reddy}, {Contini}, {Illingworth}, \& {Wilkins}}]{Atek22}
{Atek}, H., {Furtak}, L.~J., {Oesch}, P., {et~al.} 2022, \mnras, 511, 4464

\bibitem[{{Atek} {et~al.}(2023){Atek}, {Labb{\'e}}, {Furtak}, {Chemerynska}, {Fujimoto}, {Setton}, {Miller}, {Oesch}, {Bezanson}, {Price}, {Dayal}, {Zitrin}, {Kokorev}, {Weaver}, {Brammer}, {van Dokkum}, {Williams}, {Cutler}, {Feldmann}, {Fudamoto}, {Greene}, {Leja}, {Maseda}, {Muzzin}, {Pan}, {Papovich}, {Nelson}, {Nanayakkara}, {Stark}, {Stefanon}, {Suess}, {Wang}, \& {Whitaker}}]{Atek23}
{Atek}, H., {Labb{\'e}}, I., {Furtak}, L.~J., {et~al.} 2023, arXiv e-prints, arXiv:2308.08540

\bibitem[{{Bisigello} {et~al.}(2019){Bisigello}, {Caputi}, {Colina}, {P{\'e}rez-Gonz{\'a}lez}, {Koekemoer}, {Le F{\`e}vre}, {Grogin}, {N{\o}rgaard-Nielsen}, \& {van der Werf}}]{Bisigello19}
{Bisigello}, L., {Caputi}, K.~I., {Colina}, L., {et~al.} 2019, \apjs, 243, 27

\bibitem[{{B{\"o}ker} {et~al.}(2022){B{\"o}ker}, {Arribas}, {L{\"u}tzgendorf}, {Alves de Oliveira}, {Beck}, {Birkmann}, {Bunker}, {Charlot}, {de Marchi}, {Ferruit}, {Giardino}, {Jakobsen}, {Kumari}, {L{\'o}pez-Caniego}, {Maiolino}, {Manjavacas}, {Marston}, {Moseley}, {Muzerolle}, {Ogle}, {Pirzkal}, {Rauscher}, {Rawle}, {Rix}, {Sabbi}, {Sargent}, {Sirianni}, {te Plate}, {Valenti}, {Willott}, \& {Zeidler}}]{Boker22}
{B{\"o}ker}, T., {Arribas}, S., {L{\"u}tzgendorf}, N., {et~al.} 2022, \aap, 661, A82

\bibitem[{{Boucaud} {et~al.}(2016){Boucaud}, {Bocchio}, {Abergel}, {Orieux}, {Dole}, \& {Hadj-Youcef}}]{pypher}
{Boucaud}, A., {Bocchio}, M., {Abergel}, A., {et~al.} 2016, \aap, 596, A63

\bibitem[{{Bouwens} {et~al.}(2016){Bouwens}, {Smit}, {Labb{\'e}}, {Franx}, {Caruana}, {Oesch}, {Stefanon}, \& {Rasappu}}]{Bouwens16}
{Bouwens}, R.~J., {Smit}, R., {Labb{\'e}}, I., {et~al.} 2016, \apj, 831, 176

\bibitem[{{Brammer}(2019)}]{Brammer19}
{Brammer}, G. 2019, {Grizli: Grism redshift and line analysis software}, Astrophysics Source Code Library, record ascl:1905.001

\bibitem[{{Brammer} \& {Matharu}(2021)}]{grizli}
{Brammer}, G. \& {Matharu}, J. 2021, {gbrammer/grizli: Release 2021}, Zenodo

\bibitem[{{Brammer} {et~al.}(2022){Brammer}, {Strait}, {Matharu}, \& {Momcheva}}]{grizli22}
{Brammer}, G., {Strait}, V., {Matharu}, J., \& {Momcheva}, I. 2022, {grizli}, Zenodo

\bibitem[{{Bruzual} \& {Charlot}(2003)}]{BC03}
{Bruzual}, G. \& {Charlot}, S. 2003, \mnras, 344, 1000

\bibitem[{{Calzetti} {et~al.}(2000){Calzetti}, {Armus}, {Bohlin}, {Kinney}, {Koornneef}, \& {Storchi-Bergmann}}]{Calzetti00}
{Calzetti}, D., {Armus}, L., {Bohlin}, R.~C., {et~al.} 2000, \apj, 533, 682

\bibitem[{{Caputi} {et~al.}(2023){Caputi}, {Rinaldi}, {Iani}, {P{\'e}rez-Gonz{\'a}lez}, {Ostlin}, {Colina}, {Greve}, {N{\o}rgaard-Nielsen}, {Wright}, {Alvarez-M{\'a}rquez}, {Eckart}, {Hjorth}, {Labiano}, {Le F{\`e}vre}, {Walter}, {van der Werf}, {Boogaard}, {Costantin}, {Crespo-G{\'o}mez}, {Gillman}, {Jermann}, {Langeroodi}, {Melinder}, {Peissker}, \& {G{\"u}del}}]{Caputi23}
{Caputi}, K.~I., {Rinaldi}, P., {Iani}, E., {et~al.} 2023, arXiv e-prints, arXiv:2311.12691

\bibitem[{{Carnall} {et~al.}(2023){Carnall}, {Begley}, {McLeod}, {Hamadouche}, {Donnan}, {McLure}, {Dunlop}, {Milvang-Jensen}, {Bondestam}, {Cullen}, {Jewell}, \& {Pollock}}]{Carnall23}
{Carnall}, A.~C., {Begley}, R., {McLeod}, D.~J., {et~al.} 2023, \mnras, 518, L45

\bibitem[{{Carnall} {et~al.}(2019){Carnall}, {McLure}, {Dunlop}, {Cullen}, {McLeod}, {Wild}, {Johnson}, {Appleby}, {Dav{\'e}}, {Amorin}, {Bolzonella}, {Castellano}, {Cimatti}, {Cucciati}, {Gargiulo}, {Garilli}, {Marchi}, {Pentericci}, {Pozzetti}, {Schreiber}, {Talia}, \& {Zamorani}}]{Carnall19}
{Carnall}, A.~C., {McLure}, R.~J., {Dunlop}, J.~S., {et~al.} 2019, \mnras, 490, 417

\bibitem[{{Carnall} {et~al.}(2018){Carnall}, {McLure}, {Dunlop}, \& {Dav{\'e}}}]{bagpipes}
{Carnall}, A.~C., {McLure}, R.~J., {Dunlop}, J.~S., \& {Dav{\'e}}, R. 2018, \mnras, 480, 4379

\bibitem[{{Ceverino} {et~al.}(2018){Ceverino}, {Klessen}, \& {Glover}}]{Ceverino18}
{Ceverino}, D., {Klessen}, R.~S., \& {Glover}, S. C.~O. 2018, \mnras, 480, 4842

\bibitem[{{Clarke} {et~al.}(2021){Clarke}, {Scarlata}, {Mehta}, {Keel}, {Cardamone}, {Hayes}, {Adams}, {Dickinson}, {Fortson}, {Kruk}, {Lintott}, \& {Simmons}}]{Clarke21}
{Clarke}, L., {Scarlata}, C., {Mehta}, V., {et~al.} 2021, \apjl, 912, L22

\bibitem[{{Coe} {et~al.}(2019){Coe}, {Salmon}, {Brada{\v{c}}}, {Bradley}, {Sharon}, {Zitrin}, {Acebron}, {Cerny}, {Cibirka}, {Strait}, {Paterno-Mahler}, {Mahler}, {Avila}, {Ogaz}, {Huang}, {Pelliccia}, {Stark}, {Mainali}, {Oesch}, {Trenti}, {Carrasco}, {Dawson}, {Rodney}, {Strolger}, {Riess}, {Jones}, {Frye}, {Czakon}, {Umetsu}, {Vulcani}, {Graur}, {Jha}, {Graham}, {Molino}, {Nonino}, {Hjorth}, {Selsing}, {Christensen}, {Kikuchihara}, {Ouchi}, {Oguri}, {Welch}, {Lemaux}, {Andrade-Santos}, {Hoag}, {Johnson}, {Peterson}, {Past}, {Fox}, {Agulli}, {Livermore}, {Ryan}, {Lam}, {Sendra-Server}, {Toft}, {Lovisari}, \& {Su}}]{Coe19}
{Coe}, D., {Salmon}, B., {Brada{\v{c}}}, M., {et~al.} 2019, \apj, 884, 85

\bibitem[{{Conroy}(2013)}]{Conroy13}
{Conroy}, C. 2013, \araa, 51, 393

\bibitem[{{Conroy} {et~al.}(2009){Conroy}, {Gunn}, \& {White}}]{Conroy09}
{Conroy}, C., {Gunn}, J.~E., \& {White}, M. 2009, \apj, 699, 486

\bibitem[{{Conroy} {et~al.}(2010){Conroy}, {White}, \& {Gunn}}]{Conroy10}
{Conroy}, C., {White}, M., \& {Gunn}, J.~E. 2010, \apj, 708, 58

\bibitem[{{De Barros} {et~al.}(2019){De Barros}, {Oesch}, {Labb{\'e}}, {Stefanon}, {Gonz{\'a}lez}, {Smit}, {Bouwens}, \& {Illingworth}}]{DeBarros19}
{De Barros}, S., {Oesch}, P.~A., {Labb{\'e}}, I., {et~al.} 2019, \mnras, 489, 2355

\bibitem[{{Eisenstein} {et~al.}(2023){Eisenstein}, {Willott}, {Alberts}, {Arribas}, {Bonaventura}, {Bunker}, {Cameron}, {Carniani}, {Charlot}, {Curtis-Lake}, {D'Eugenio}, {Endsley}, {Ferruit}, {Giardino}, {Hainline}, {Hausen}, {Jakobsen}, {Johnson}, {Maiolino}, {Rieke}, {Rieke}, {Rix}, {Robertson}, {Stark}, {Tacchella}, {Williams}, {Willmer}, {Baker}, {Baum}, {Bhatawdekar}, {Boyett}, {Chen}, {Chevallard}, {Circosta}, {Curti}, {Danhaive}, {DeCoursey}, {de Graaff}, {Dressler}, {Egami}, {Helton}, {Hviding}, {Ji}, {Jones}, {Kumari}, {L{\"u}tzgendorf}, {Laseter}, {Looser}, {Lyu}, {Maseda}, {Nelson}, {Parlanti}, {Perna}, {Pusk{\'a}s}, {Rawle}, {Rodr{\'i}guez Del Pino}, {Sandles}, {Saxena}, {Scholtz}, {Sharpe}, {Shivaei}, {Silcock}, {Simmonds}, {Skarbinski}, {Smit}, {Stone}, {Suess}, {Sun}, {Tang}, {Topping}, {{\"U}bler}, {Villanueva}, {Wallace}, {Whitler}, {Witstok}, \& {Woodrum}}]{Eisenstein23}
{Eisenstein}, D.~J., {Willott}, C., {Alberts}, S., {et~al.} 2023, arXiv e-prints, arXiv:2306.02465

\bibitem[{{Endsley} {et~al.}(2021){Endsley}, {Stark}, {Chevallard}, \& {Charlot}}]{Endsley21}
{Endsley}, R., {Stark}, D.~P., {Chevallard}, J., \& {Charlot}, S. 2021, \mnras, 500, 5229

\bibitem[{{Endsley} {et~al.}(2023){Endsley}, {Stark}, {Whitler}, {Topping}, {Chen}, {Plat}, {Chisholm}, \& {Charlot}}]{Endsley23}
{Endsley}, R., {Stark}, D.~P., {Whitler}, L., {et~al.} 2023, \mnras, 524, 2312

\bibitem[{{Fan} {et~al.}(2006){Fan}, {Strauss}, {Becker}, {White}, {Gunn}, {Knapp}, {Richards}, {Schneider}, {Brinkmann}, \& {Fukugita}}]{Fan06}
{Fan}, X., {Strauss}, M.~A., {Becker}, R.~H., {et~al.} 2006, \aj, 132, 117

\bibitem[{{Ferland} {et~al.}(2017){Ferland}, {Chatzikos}, {Guzm{\'a}n}, {Lykins}, {van Hoof}, {Williams}, {Abel}, {Badnell}, {Keenan}, {Porter}, \& {Stancil}}]{cloudy}
{Ferland}, G.~J., {Chatzikos}, M., {Guzm{\'a}n}, F., {et~al.} 2017, \rmxaa, 53, 385

\bibitem[{{Fitzpatrick} \& {Massa}(2007)}]{Fitzpatrick07}
{Fitzpatrick}, E.~L. \& {Massa}, D. 2007, \apj, 663, 320

\bibitem[{{Fujimoto} {et~al.}(2021{\natexlab{a}}){Fujimoto}, {Ao}, {Bartosch Caminha}, {Bauer}, {Brammer}, {Dessauges-Zavadsky}, {Egami}, {Fudamoto}, {Hatsukade}, {Hughes}, {Knudsen}, {Koekemoer}, {Kohno}, {Laporte}, {Lee}, {Magdis}, {Oguri}, {Ouchi}, {Rawle}, {Richard}, {Rujopakarn}, {Shimasaku}, {Sun}, {Umehata}, {Urbina}, {Valentino}, {Wang}, {Wang}, \& {Zitrin}}]{2021jwst.prop.1567F}
{Fujimoto}, S., {Ao}, Y., {Bartosch Caminha}, G., {et~al.} 2021{\natexlab{a}}, {Early Galaxy Assembly Uncovered with ALMA and JWST: A Remarkably UV and [CII] Bright, Strongly Lensed Sub-L* Galaxy at z=6.072}, JWST Proposal. Cycle 1, ID. \#1567

\bibitem[{{Fujimoto} {et~al.}(2023){Fujimoto}, {Arrabal Haro}, {Dickinson}, {Finkelstein}, {Kartaltepe}, {Larson}, {Burgarella}, {Bagley}, {Behroozi}, {Chworowsky}, {Hirschmann}, {Trump}, {Wilkins}, {Yung}, {Koekemoer}, {Papovich}, {Pirzkal}, {Ferguson}, {Fontana}, {Grogin}, {Grazian}, {Kewley}, {Kocevski}, {Lotz}, {Pentericci}, {Ravindranath}, {Somerville}, {Wilkins}, {Amor{\'i}n}, {Backhaus}, {Calabr{\`o}}, {Casey}, {Cooper}, {Fern{\'a}ndez}, {Franco}, {Giavalisco}, {Hathi}, {Harish}, {Hutchison}, {Iyer}, {Jung}, {Lucas}, \& {Zavala}}]{Fujimoto23}
{Fujimoto}, S., {Arrabal Haro}, P., {Dickinson}, M., {et~al.} 2023, \apjl, 949, L25

\bibitem[{{Fujimoto} {et~al.}(2021{\natexlab{b}}){Fujimoto}, {Oguri}, {Brammer}, {Yoshimura}, {Laporte}, {Gonz{\'a}lez-L{\'o}pez}, {Caminha}, {Kohno}, {Zitrin}, {Richard}, {Ouchi}, {Bauer}, {Smail}, {Hatsukade}, {Ono}, {Kokorev}, {Umehata}, {Schaerer}, {Knudsen}, {Sun}, {Magdis}, {Valentino}, {Ao}, {Toft}, {Dessauges-Zavadsky}, {Shimasaku}, {Caputi}, {Kusakabe}, {Morokuma-Matsui}, {Shotaro}, {Egami}, {Lee}, {Rawle}, \& {Espada}}]{Fujimoto21}
{Fujimoto}, S., {Oguri}, M., {Brammer}, G., {et~al.} 2021{\natexlab{b}}, \apj, 911, 99

\bibitem[{{Gim{\'e}nez-Arteaga} {et~al.}(2022){Gim{\'e}nez-Arteaga}, {Brammer}, {Marchesini}, {Colina}, {Bajaj}, {Brinch}, {Calzetti}, {Lange-Vagle}, {Murphy}, {Perna}, {Piqueras-L{\'o}pez}, \& {Snyder}}]{Arteaga22}
{Gim{\'e}nez-Arteaga}, C., {Brammer}, G.~B., {Marchesini}, D., {et~al.} 2022, \apjs, 263, 17

\bibitem[{{Gim{\'e}nez-Arteaga} {et~al.}(2023){Gim{\'e}nez-Arteaga}, {Oesch}, {Brammer}, {Valentino}, {Mason}, {Weibel}, {Barrufet}, {Fujimoto}, {Heintz}, {Nelson}, {Strait}, {Suess}, \& {Gibson}}]{Arteaga23}
{Gim{\'e}nez-Arteaga}, C., {Oesch}, P.~A., {Brammer}, G.~B., {et~al.} 2023, \apj, 948, 126

\bibitem[{{Graves} \& {Faber}(2010)}]{Graves10}
{Graves}, G.~J. \& {Faber}, S.~M. 2010, \apj, 717, 803

\bibitem[{Harris {et~al.}(2020)Harris, Millman, van~der Walt, Gommers, Virtanen, Cournapeau, Wieser, Taylor, Berg, Smith, Kern, Picus, Hoyer, van Kerkwijk, Brett, Haldane, del R{\'{i}}o, Wiebe, Peterson, G{\'{e}}rard-Marchant, Sheppard, Reddy, Weckesser, Abbasi, Gohlke, \& Oliphant}]{numpy}
Harris, C.~R., Millman, K.~J., van~der Walt, S.~J., {et~al.} 2020, Nature, 585, 357

\bibitem[{{Hashimoto} {et~al.}(2018){Hashimoto}, {Laporte}, {Mawatari}, {Ellis}, {Inoue}, {Zackrisson}, {Roberts-Borsani}, {Zheng}, {Tamura}, {Bauer}, {Fletcher}, {Harikane}, {Hatsukade}, {Hayatsu}, {Matsuda}, {Matsuo}, {Okamoto}, {Ouchi}, {Pell{\'o}}, {Rydberg}, {Shimizu}, {Taniguchi}, {Umehata}, \& {Yoshida}}]{Hashimoto18}
{Hashimoto}, T., {Laporte}, N., {Mawatari}, K., {et~al.} 2018, \nat, 557, 392

\bibitem[{{Heintz} {et~al.}(2023){Heintz}, {Gim{\'e}nez-Arteaga}, {Fujimoto}, {Brammer}, {Espada}, {Gillman}, {Gonz{\'a}lez-L{\'o}pez}, {Greve}, {Harikane}, {Hatsukade}, {Knudsen}, {Koekemoer}, {Kohno}, {Kokorev}, {Lee}, {Magdis}, {Nelson}, {Rizzo}, {Sanders}, {Schaerer}, {Shapley}, {Strait}, {Toft}, {Valentino}, {van der Wel}, {Vijayan}, {Watson}, {Bauer}, {Christiansen}, \& {Wilson}}]{Heintz23}
{Heintz}, K.~E., {Gim{\'e}nez-Arteaga}, C., {Fujimoto}, S., {et~al.} 2023, \apjl, 944, L30

\bibitem[{Hunter(2007)}]{Hunter:2007}
Hunter, J.~D. 2007, Computing in Science \& Engineering, 9, 90

\bibitem[{{Illingworth} {et~al.}(2013){Illingworth}, {Magee}, {Oesch}, {Bouwens}, {Labb{\'e}}, {Stiavelli}, {van Dokkum}, {Franx}, {Trenti}, {Carollo}, \& {Gonzalez}}]{xdf}
{Illingworth}, G.~D., {Magee}, D., {Oesch}, P.~A., {et~al.} 2013, \apjs, 209, 6

\bibitem[{{Inoue} {et~al.}(2014){Inoue}, {Shimizu}, {Iwata}, \& {Tanaka}}]{Inoue14}
{Inoue}, A.~K., {Shimizu}, I., {Iwata}, I., \& {Tanaka}, M. 2014, \mnras, 442, 1805

\bibitem[{{Iyer} {et~al.}(2020){Iyer}, {Tacchella}, {Genel}, {Hayward}, {Hernquist}, {Brooks}, {Caplar}, {Dav{\'e}}, {Diemer}, {Forbes}, {Gawiser}, {Somerville}, \& {Starkenburg}}]{Iyer20}
{Iyer}, K.~G., {Tacchella}, S., {Genel}, S., {et~al.} 2020, \mnras, 498, 430

\bibitem[{{Jain} {et~al.}(2024){Jain}, {Tacchella}, \& {Mosleh}}]{Jain24}
{Jain}, S., {Tacchella}, S., \& {Mosleh}, M. 2024, \mnras, 527, 3291

\bibitem[{{Kroupa}(2001)}]{Kroupa01}
{Kroupa}, P. 2001, \mnras, 322, 231

\bibitem[{{Labb{\'e}} {et~al.}(2013){Labb{\'e}}, {Oesch}, {Bouwens}, {Illingworth}, {Magee}, {Gonz{\'a}lez}, {Carollo}, {Franx}, {Trenti}, {van Dokkum}, \& {Stiavelli}}]{Labbe13}
{Labb{\'e}}, I., {Oesch}, P.~A., {Bouwens}, R.~J., {et~al.} 2013, \apjl, 777, L19

\bibitem[{{Labb{\'e}} {et~al.}(2023){Labb{\'e}}, {van Dokkum}, {Nelson}, {Bezanson}, {Suess}, {Leja}, {Brammer}, {Whitaker}, {Mathews}, {Stefanon}, \& {Wang}}]{Labbe23}
{Labb{\'e}}, I., {van Dokkum}, P., {Nelson}, E., {et~al.} 2023, \nat, 616, 266

\bibitem[{{Lam} {et~al.}(2019){Lam}, {Bouwens}, {Labb{\'e}}, {Schaye}, {Schmidt}, {Maseda}, {Bacon}, {Boogaard}, {Nanayakkara}, {Richard}, {Mahler}, \& {Urrutia}}]{Lam19}
{Lam}, D., {Bouwens}, R.~J., {Labb{\'e}}, I., {et~al.} 2019, \aap, 627, A164

\bibitem[{{Laporte} {et~al.}(2021){Laporte}, {Zitrin}, {Ellis}, {Fujimoto}, {Brammer}, {Richard}, {Oguri}, {Caminha}, {Kohno}, {Yoshimura}, {Ao}, {Bauer}, {Caputi}, {Egami}, {Espada}, {Gonz{\'a}lez-L{\'o}pez}, {Hatsukade}, {Knudsen}, {Lee}, {Magdis}, {Ouchi}, {Valentino}, \& {Wang}}]{Laporte21}
{Laporte}, N., {Zitrin}, A., {Ellis}, R.~S., {et~al.} 2021, \mnras, 505, 4838

\bibitem[{{Leja} {et~al.}(2019){Leja}, {Carnall}, {Johnson}, {Conroy}, \& {Speagle}}]{Leja19}
{Leja}, J., {Carnall}, A.~C., {Johnson}, B.~D., {Conroy}, C., \& {Speagle}, J.~S. 2019, \apj, 876, 3

\bibitem[{{Leja} {et~al.}(2017){Leja}, {Johnson}, {Conroy}, {van Dokkum}, \& {Byler}}]{Leja17}
{Leja}, J., {Johnson}, B.~D., {Conroy}, C., {van Dokkum}, P.~G., \& {Byler}, N. 2017, \apj, 837, 170

\bibitem[{{Lupton} {et~al.}(2004){Lupton}, {Blanton}, {Fekete}, {Hogg}, {O'Mullane}, {Szalay}, \& {Wherry}}]{Lupton04}
{Lupton}, R., {Blanton}, M.~R., {Fekete}, G., {et~al.} 2004, \pasp, 116, 133

\bibitem[{{Maiolino} {et~al.}(2023){Maiolino}, {Uebler}, {Perna}, {Scholtz}, {D'Eugenio}, {Witten}, {Laporte}, {Witstok}, {Carniani}, {Tacchella}, {Baker}, {Arribas}, {Nakajima}, {Eisenstein}, {Bunker}, {Charlot}, {Cresci}, {Curti}, {Curtis-Lake}, {de Graaff}, {Ji}, {Johnson}, {Kumari}, {Looser}, {Maseda}, {Robertson}, {Rodriguez Del Pino}, {Sandles}, {Simmonds}, {Smit}, {Sun}, {Venturi}, {Williams}, \& {Willmer}}]{Maiolino23}
{Maiolino}, R., {Uebler}, H., {Perna}, M., {et~al.} 2023, arXiv e-prints, arXiv:2306.00953

\bibitem[{{Maraston} {et~al.}(2013){Maraston}, {Pforr}, {Henriques}, {Thomas}, {Wake}, {Brownstein}, {Capozzi}, {Tinker}, {Bundy}, {Skibba}, {Beifiori}, {Nichol}, {Edmondson}, {Schneider}, {Chen}, {Masters}, {Steele}, {Bolton}, {York}, {Weaver}, {Higgs}, {Bizyaev}, {Brewington}, {Malanushenko}, {Malanushenko}, {Snedden}, {Oravetz}, {Pan}, {Shelden}, \& {Simmons}}]{Maraston13}
{Maraston}, C., {Pforr}, J., {Henriques}, B.~M., {et~al.} 2013, \mnras, 435, 2764

\bibitem[{{Maraston} {et~al.}(2010){Maraston}, {Pforr}, {Renzini}, {Daddi}, {Dickinson}, {Cimatti}, \& {Tonini}}]{Maraston10}
{Maraston}, C., {Pforr}, J., {Renzini}, A., {et~al.} 2010, \mnras, 407, 830

\bibitem[{{Marsan} {et~al.}(2022){Marsan}, {Muzzin}, {Marchesini}, {Stefanon}, {Martis}, {Annunziatella}, {Chan}, {Cooper}, {Forrest}, {Gomez}, {McConachie}, \& {Wilson}}]{Marsan22}
{Marsan}, Z.~C., {Muzzin}, A., {Marchesini}, D., {et~al.} 2022, \apj, 924, 25

\bibitem[{{Mason} {et~al.}(2018){Mason}, {Treu}, {Dijkstra}, {Mesinger}, {Trenti}, {Pentericci}, {de Barros}, \& {Vanzella}}]{Mason18}
{Mason}, C.~A., {Treu}, T., {Dijkstra}, M., {et~al.} 2018, \apj, 856, 2

\bibitem[{{Matthee} {et~al.}(2023){Matthee}, {Mackenzie}, {Simcoe}, {Kashino}, {Lilly}, {Bordoloi}, \& {Eilers}}]{Matthee23}
{Matthee}, J., {Mackenzie}, R., {Simcoe}, R.~A., {et~al.} 2023, \apj, 950, 67

\bibitem[{{Matthee} {et~al.}(2017){Matthee}, {Sobral}, {Best}, {Khostovan}, {Oteo}, {Bouwens}, \& {R{\"o}ttgering}}]{Matthee17}
{Matthee}, J., {Sobral}, D., {Best}, P., {et~al.} 2017, \mnras, 465, 3637

\bibitem[{{Narayanan} {et~al.}(2023){Narayanan}, {Lower}, {Torrey}, {Brammer}, {Cui}, {Dave}, {Iyer}, {Li}, {Lovell}, {Sales}, {Stark}, {Marinacci}, \& {Vogelsberger}}]{Narayanan23}
{Narayanan}, D., {Lower}, S., {Torrey}, P., {et~al.} 2023, arXiv e-prints, arXiv:2306.10118

\bibitem[{{Pacifici} {et~al.}(2023){Pacifici}, {Iyer}, {Mobasher}, {da Cunha}, {Acquaviva}, {Burgarella}, {Calistro Rivera}, {Carnall}, {Chang}, {Chartab}, {Cooke}, {Fairhurst}, {Kartaltepe}, {Leja}, {Ma{\l}ek}, {Salmon}, {Torelli}, {Vidal-Garc{\'i}a}, {Boquien}, {Brammer}, {Brown}, {Capak}, {Chevallard}, {Circosta}, {Croton}, {Davidzon}, {Dickinson}, {Duncan}, {Faber}, {Ferguson}, {Fontana}, {Guo}, {Haeussler}, {Hemmati}, {Jafariyazani}, {Kassin}, {Larson}, {Lee}, {Mantha}, {Marchi}, {Nayyeri}, {Newman}, {Pandya}, {Pforr}, {Reddy}, {Sanders}, {Shah}, {Shahidi}, {Stevans}, {Triani}, {Tyler}, {Vanderhoof}, {de la Vega}, {Wang}, \& {Weston}}]{Pacifici23}
{Pacifici}, C., {Iyer}, K.~G., {Mobasher}, B., {et~al.} 2023, \apj, 944, 141

\bibitem[{{Papovich} {et~al.}(2001){Papovich}, {Dickinson}, \& {Ferguson}}]{Papovich01}
{Papovich}, C., {Dickinson}, M., \& {Ferguson}, H.~C. 2001, \apj, 559, 620

\bibitem[{{Paulino-Afonso} {et~al.}(2022){Paulino-Afonso}, {Gonz{\'a}lez-Gait{\'a}n}, {Galbany}, {Maria Mour{\~a}o}, {Angus}, {Smith}, {Anderson}, {Lyman}, {Kuncarayakti}, \& {Rodrigues}}]{PaulinoAfonso22}
{Paulino-Afonso}, A., {Gonz{\'a}lez-Gait{\'a}n}, S., {Galbany}, L., {et~al.} 2022, \aap, 662, A86

\bibitem[{{Perrin} {et~al.}(2014){Perrin}, {Sivaramakrishnan}, {Lajoie}, {Elliott}, {Pueyo}, {Ravindranath}, \& {Albert}}]{Perrin14}
{Perrin}, M.~D., {Sivaramakrishnan}, A., {Lajoie}, C.-P., {et~al.} 2014, in Society of Photo-Optical Instrumentation Engineers (SPIE) Conference Series, Vol. 9143, Space Telescopes and Instrumentation 2014: Optical, Infrared, and Millimeter Wave, ed. J.~{Oschmann}, Jacobus~M., M.~{Clampin}, G.~G. {Fazio}, \& H.~A. {MacEwen}, 91433X

\bibitem[{{Perrin} {et~al.}(2012){Perrin}, {Soummer}, {Elliott}, {Lallo}, \& {Sivaramakrishnan}}]{Perrin12}
{Perrin}, M.~D., {Soummer}, R., {Elliott}, E.~M., {Lallo}, M.~D., \& {Sivaramakrishnan}, A. 2012, in Society of Photo-Optical Instrumentation Engineers (SPIE) Conference Series, Vol. 8442, Space Telescopes and Instrumentation 2012: Optical, Infrared, and Millimeter Wave, ed. M.~C. {Clampin}, G.~G. {Fazio}, H.~A. {MacEwen}, \& J.~{Oschmann}, Jacobus~M., 84423D

\bibitem[{{Pforr} {et~al.}(2012){Pforr}, {Maraston}, \& {Tonini}}]{Pforr12}
{Pforr}, J., {Maraston}, C., \& {Tonini}, C. 2012, \mnras, 422, 3285

\bibitem[{{Pforr} {et~al.}(2013){Pforr}, {Maraston}, \& {Tonini}}]{Pforr13}
{Pforr}, J., {Maraston}, C., \& {Tonini}, C. 2013, \mnras, 435, 1389

\bibitem[{{Prieto-Lyon} {et~al.}(2023){Prieto-Lyon}, {Strait}, {Mason}, {Brammer}, {Caminha}, {Mercurio}, {Acebron}, {Bergamini}, {Grillo}, {Rosati}, {Vanzella}, {Castellano}, {Merlin}, {Paris}, {Boyett}, {Calabr{\`o}}, {Morishita}, {Mascia}, {Pentericci}, {Roberts-Borsani}, {Roy}, {Treu}, \& {Vulcani}}]{Prieto-Lyon23}
{Prieto-Lyon}, G., {Strait}, V., {Mason}, C.~A., {et~al.} 2023, \aap, 672, A186

\bibitem[{{Rieke} {et~al.}(2005){Rieke}, {Kelly}, \& {Horner}}]{2005SPIE.5904....1R}
{Rieke}, M.~J., {Kelly}, D., \& {Horner}, S. 2005, in Society of Photo-Optical Instrumentation Engineers (SPIE) Conference Series, Vol. 5904, Cryogenic Optical Systems and Instruments XI, ed. J.~B. {Heaney} \& L.~G. {Burriesci}, 1--8

\bibitem[{{Rieke} {et~al.}(2023){Rieke}, {Kelly}, {Misselt}, {Stansberry}, {Boyer}, {Beatty}, {Egami}, {Florian}, {Greene}, {Hainline}, {Leisenring}, {Roellig}, {Schlawin}, {Sun}, {Tinnin}, {Williams}, {Willmer}, {Wilson}, {Clark}, {Rohrbach}, {Brooks}, {Canipe}, {Correnti}, {DiFelice}, {Gennaro}, {Girard}, {Hartig}, {Hilbert}, {Koekemoer}, {Nikolov}, {Pirzkal}, {Rest}, {Robberto}, {Sunnquist}, {Telfer}, {Wu}, {Ferry}, {Lewis}, {Baum}, {Beichman}, {Doyon}, {Dressler}, {Eisenstein}, {Ferrarese}, {Hodapp}, {Horner}, {Jaffe}, {Johnstone}, {Krist}, {Martin}, {McCarthy}, {Meyer}, {Rieke}, {Trauger}, \& {Young}}]{Rieke23}
{Rieke}, M.~J., {Kelly}, D.~M., {Misselt}, K., {et~al.} 2023, \pasp, 135, 028001

\bibitem[{{Rigby} {et~al.}(2023){Rigby}, {Vieira}, {Phadke}, {Hutchison}, {Welch}, {Cathey}, {Spilker}, {Gonzalez}, {Adhikari}, {Aravena}, {Bayliss}, {Birkin}, {Bursk}, {Chapman}, {Dahle}, {Elicker}, {Fischer}, {Florian}, {Gladders}, {Hayward}, {Hewald}, {Kettler}, {Khullar}, {Kim}, {Law}, {Mahler}, {Malhotra}, {Murphy}, {Narayanan}, {Olivier}, {Rhoads}, {Sharon}, {Solimano}, {Thiruvengadam}, {Vizgan}, \& {Younker}}]{Rigby23}
{Rigby}, J.~R., {Vieira}, J.~D., {Phadke}, K.~A., {et~al.} 2023, arXiv e-prints, arXiv:2312.10465

\bibitem[{{Rinaldi} {et~al.}(2023){Rinaldi}, {Caputi}, {Costantin}, {Gillman}, {Iani}, {P{\'e}rez-Gonz{\'a}lez}, {{\"O}stlin}, {Colina}, {Greve}, {Noorgard-Nielsen}, {Wright}, {Alonso-Herrero}, {{\'A}lvarez-M{\'a}rquez}, {Eckart}, {Garc{\'i}a-Mar{\'i}n}, {Hjorth}, {Ilbert}, {Kendrew}, {Labiano}, {Le F{\`e}vre}, {Pye}, {Tikkanen}, {Walter}, {van der Werf}, {Ward}, {Annunziatella}, {Azzollini}, {Bik}, {Boogaard}, {Bosman}, {Crespo G{\'o}mez}, {Jermann}, {Langeroodi}, {Melinder}, {Meyer}, {Moutard}, {Peissker}, {Topinka}, {van Dishoeck}, {G{\"u}del}, {Henning}, {Lagage}, {Ray}, {Vandenbussche}, {Waelkens}, {Navarro-Carrera}, \& {Kokorev}}]{Rinaldi23}
{Rinaldi}, P., {Caputi}, K.~I., {Costantin}, L., {et~al.} 2023, \apj, 952, 143

\bibitem[{{Roberts-Borsani} {et~al.}(2020){Roberts-Borsani}, {Ellis}, \& {Laporte}}]{Borsani20}
{Roberts-Borsani}, G.~W., {Ellis}, R.~S., \& {Laporte}, N. 2020, \mnras, 497, 3440

\bibitem[{{Robertson} {et~al.}(2013){Robertson}, {Furlanetto}, {Schneider}, {Charlot}, {Ellis}, {Stark}, {McLure}, {Dunlop}, {Koekemoer}, {Schenker}, {Ouchi}, {Ono}, {Curtis-Lake}, {Rogers}, {Bowler}, \& {Cirasuolo}}]{Robertson13}
{Robertson}, B.~E., {Furlanetto}, S.~R., {Schneider}, E., {et~al.} 2013, \apj, 768, 71

\bibitem[{{Rusakov} {et~al.}(2023){Rusakov}, {Steinhardt}, \& {Sneppen}}]{Rusakov23}
{Rusakov}, V., {Steinhardt}, C.~L., \& {Sneppen}, A. 2023, \apjs, 268, 10

\bibitem[{{Sawicki} \& {Yee}(1998)}]{Sawicki98}
{Sawicki}, M. \& {Yee}, H.~K.~C. 1998, \aj, 115, 1329

\bibitem[{{Schlafly} \& {Finkbeiner}(2011)}]{MW11}
{Schlafly}, E.~F. \& {Finkbeiner}, D.~P. 2011, \apj, 737, 103

\bibitem[{{Shapley} {et~al.}(2001){Shapley}, {Steidel}, {Adelberger}, {Dickinson}, {Giavalisco}, \& {Pettini}}]{Shapley01}
{Shapley}, A.~E., {Steidel}, C.~C., {Adelberger}, K.~L., {et~al.} 2001, \apj, 562, 95

\bibitem[{{Shivaei} {et~al.}(2018){Shivaei}, {Reddy}, {Siana}, {Shapley}, {Kriek}, {Mobasher}, {Freeman}, {Sanders}, {Coil}, {Price}, {Fetherolf}, {Azadi}, {Leung}, \& {Zick}}]{Shivaei18}
{Shivaei}, I., {Reddy}, N.~A., {Siana}, B., {et~al.} 2018, \apj, 855, 42

\bibitem[{{Simmonds} {et~al.}(2023){Simmonds}, {Tacchella}, {Hainline}, {Johnson}, {McClymont}, {Robertson}, {Saxena}, {Sun}, {Witten}, {Baker}, {Bhatawdekar}, {Boyett}, {Bunker}, {Charlot}, {Curtis-Lake}, {Egami}, {Eisenstein}, {Hausen}, {Maiolino}, {Maseda}, {Scholtz}, {Williams}, {Willot}, \& {Witstok}}]{Simmonds23}
{Simmonds}, C., {Tacchella}, S., {Hainline}, K., {et~al.} 2023, arXiv e-prints, arXiv:2310.01112

\bibitem[{{Smail} {et~al.}(2023){Smail}, {Dudzevi{\v{c}}i{\={u}}t{\.{e}}}, {Gurwell}, {Fazio}, {Willner}, {Swinbank}, {Arumugam}, {Summers}, {Cohen}, {Jansen}, {Windhorst}, {Meena}, {Zitrin}, {Keel}, {Cheng}, {Coe}, {Conselice}, {D'Silva}, {Driver}, {Frye}, {Grogin}, {Koekemoer}, {Marshall}, {Nonino}, {Pirzkal}, {Robotham}, {Rutkowski}, {Ryan}, {Tompkins}, {Willmer}, {Yan}, {Broadhurst}, {Diego}, {Kamieneski}, \& {Yun}}]{Smail23}
{Smail}, I., {Dudzevi{\v{c}}i{\={u}}t{\.{e}}}, U., {Gurwell}, M., {et~al.} 2023, \apj, 958, 36

\bibitem[{{Smit} {et~al.}(2016){Smit}, {Bouwens}, {Labb{\'e}}, {Franx}, {Wilkins}, \& {Oesch}}]{Smit16}
{Smit}, R., {Bouwens}, R.~J., {Labb{\'e}}, I., {et~al.} 2016, \apj, 833, 254

\bibitem[{{Smit} {et~al.}(2014){Smit}, {Bouwens}, {Labb{\'e}}, {Zheng}, {Bradley}, {Donahue}, {Lemze}, {Moustakas}, {Umetsu}, {Zitrin}, {Coe}, {Postman}, {Gonzalez}, {Bartelmann}, {Ben{\'i}tez}, {Broadhurst}, {Ford}, {Grillo}, {Infante}, {Jimenez-Teja}, {Jouvel}, {Kelson}, {Lahav}, {Maoz}, {Medezinski}, {Melchior}, {Meneghetti}, {Merten}, {Molino}, {Moustakas}, {Nonino}, {Rosati}, \& {Seitz}}]{Smit14}
{Smit}, R., {Bouwens}, R.~J., {Labb{\'e}}, I., {et~al.} 2014, \apj, 784, 58

\bibitem[{{Song} {et~al.}(2023){Song}, {Fang}, {Lin}, {Gu}, \& {Kong}}]{Song23}
{Song}, J., {Fang}, G., {Lin}, Z., {Gu}, Y., \& {Kong}, X. 2023, \apj, 958, 82

\bibitem[{{Sorba} \& {Sawicki}(2015)}]{Sorba15}
{Sorba}, R. \& {Sawicki}, M. 2015, \mnras, 452, 235

\bibitem[{{Sorba} \& {Sawicki}(2018)}]{Sorba18}
{Sorba}, R. \& {Sawicki}, M. 2018, \mnras, 476, 1532

\bibitem[{{Stark} {et~al.}(2013){Stark}, {Schenker}, {Ellis}, {Robertson}, {McLure}, \& {Dunlop}}]{Stark13}
{Stark}, D.~P., {Schenker}, M.~A., {Ellis}, R., {et~al.} 2013, \apj, 763, 129

\bibitem[{{Stefanon} {et~al.}(2022){Stefanon}, {Bouwens}, {Illingworth}, {Labb{\'e}}, {Oesch}, \& {Gonzalez}}]{Stefanon22}
{Stefanon}, M., {Bouwens}, R.~J., {Illingworth}, G.~D., {et~al.} 2022, \apj, 935, 94

\bibitem[{{Steinhardt} {et~al.}(2023){Steinhardt}, {Kokorev}, {Rusakov}, {Garcia}, \& {Sneppen}}]{Steinhardt23}
{Steinhardt}, C.~L., {Kokorev}, V., {Rusakov}, V., {Garcia}, E., \& {Sneppen}, A. 2023, \apjl, 951, L40

\bibitem[{{Strait} {et~al.}(2021){Strait}, {Brada{\v{c}}}, {Coe}, {Lemaux}, {Carnall}, {Bradley}, {Pelliccia}, {Sharon}, {Zitrin}, {Acebron}, {Neufeld}, {Andrade-Santos}, {Avila}, {Frye}, {Mahler}, {Nonino}, {Ogaz}, {Oguri}, {Ouchi}, {Paterno-Mahler}, {Stark}, {Mainali}, {Oesch}, {Trenti}, {Carrasco}, {Dawson}, {Jones}, {Umetsu}, \& {Vulcani}}]{Strait21}
{Strait}, V., {Brada{\v{c}}}, M., {Coe}, D., {et~al.} 2021, \apj, 910, 135

\bibitem[{{Strait} {et~al.}(2023){Strait}, {Brammer}, {Muzzin}, {Desprez}, {Asada}, {Abraham}, {Brada{\v{c}}}, {Iyer}, {Martis}, {Mowla}, {Noirot}, {Sarrouh}, {Sawicki}, {Willott}, {Gould}, {Grindlay}, {Matharu}, \& {Rihtar{\v{s}}i{\v{c}}}}]{Strait23}
{Strait}, V., {Brammer}, G., {Muzzin}, A., {et~al.} 2023, \apjl, 949, L23

\bibitem[{{Suess} {et~al.}(2022){Suess}, {Leja}, {Johnson}, {Bezanson}, {Greene}, {Kriek}, {Lower}, {Narayanan}, {Setton}, \& {Spilker}}]{Suess22}
{Suess}, K.~A., {Leja}, J., {Johnson}, B.~D., {et~al.} 2022, \apj, 935, 146

\bibitem[{{Tacchella} {et~al.}(2023){Tacchella}, {Johnson}, {Robertson}, {Carniani}, {D'Eugenio}, {Kumari}, {Maiolino}, {Nelson}, {Suess}, {{\"U}bler}, {Williams}, {Adebusola}, {Alberts}, {Arribas}, {Bhatawdekar}, {Bonaventura}, {Bowler}, {Bunker}, {Cameron}, {Curti}, {Egami}, {Eisenstein}, {Frye}, {Hainline}, {Helton}, {Ji}, {Looser}, {Lyu}, {Perna}, {Rawle}, {Rieke}, {Rieke}, {Saxena}, {Sandles}, {Shivaei}, {Simmonds}, {Sun}, {Willmer}, {Willott}, \& {Witstok}}]{Tacchella23}
{Tacchella}, S., {Johnson}, B.~D., {Robertson}, B.~E., {et~al.} 2023, \mnras, 522, 6236

\bibitem[{{Tang} {et~al.}(2022){Tang}, {Stark}, \& {Ellis}}]{Tang22}
{Tang}, M., {Stark}, D.~P., \& {Ellis}, R.~S. 2022, \mnras, 513, 5211

\bibitem[{{Topping} {et~al.}(2022){Topping}, {Stark}, {Endsley}, {Bouwens}, {Schouws}, {Smit}, {Stefanon}, {Inami}, {Bowler}, {Oesch}, {Gonzalez}, {Dayal}, {da Cunha}, {Algera}, {van der Werf}, {Pallottini}, {Barrufet}, {Schneider}, {De Looze}, {Sommovigo}, {Whitler}, {Graziani}, {Fudamoto}, \& {Ferrara}}]{Topping22}
{Topping}, M.~W., {Stark}, D.~P., {Endsley}, R., {et~al.} 2022, \mnras, 516, 975

\bibitem[{{Trager} {et~al.}(2008){Trager}, {Faber}, \& {Dressler}}]{Trager08}
{Trager}, S.~C., {Faber}, S.~M., \& {Dressler}, A. 2008, \mnras, 386, 715

\bibitem[{{Valentino} {et~al.}(2023){Valentino}, {Brammer}, {Gould}, {Kokorev}, {Fujimoto}, {Jespersen}, {Vijayan}, {Weaver}, {Ito}, {Tanaka}, {Ilbert}, {Magdis}, {Whitaker}, {Faisst}, {Gallazzi}, {Gillman}, {Gim{\'e}nez-Arteaga}, {G{\'o}mez-Guijarro}, {Kubo}, {Heintz}, {Hirschmann}, {Oesch}, {Onodera}, {Rizzo}, {Lee}, {Strait}, \& {Toft}}]{Valentino23}
{Valentino}, F., {Brammer}, G., {Gould}, K. M.~L., {et~al.} 2023, \apj, 947, 20

\bibitem[{{Valentino} {et~al.}(2024){Valentino}, {Fujimoto}, {Gim{\'e}nez-Arteaga}, {Brammer}, {Kohno}, \& {Sun}}]{Valentino24}
{Valentino}, F., {Fujimoto}, S., {Gim{\'e}nez-Arteaga}, C., {et~al.} 2024, \aap, submitted

\bibitem[{Virtanen {et~al.}(2020)Virtanen, Gommers, Oliphant, Haberland, Reddy, Cournapeau, Burovski, Peterson, Weckesser, Bright, {van der Walt}, Brett, Wilson, Millman, Mayorov, Nelson, Jones, Kern, Larson, Carey, Polat, Feng, Moore, {VanderPlas}, Laxalde, Perktold, Cimrman, Henriksen, Quintero, Harris, Archibald, Ribeiro, Pedregosa, {van Mulbregt}, \& {SciPy 1.0 Contributors}}]{scipy}
Virtanen, P., Gommers, R., Oliphant, T.~E., {et~al.} 2020, Nature Methods, 17, 261

\bibitem[{{Wang} {et~al.}(2023){Wang}, {Leja}, {Atek}, {Labbe}, {Li}, {Bezanson}, {Brammer}, {Cutler}, {Dayal}, {Furtak}, {Greene}, {Kokorev}, {Pan}, {Price}, {Suess}, {Weaver}, {Whitaker}, \& {Williams}}]{Wang23}
{Wang}, B., {Leja}, J., {Atek}, H., {et~al.} 2023, arXiv e-prints, arXiv:2310.06781

\bibitem[{{Whitler} {et~al.}(2023){Whitler}, {Stark}, {Endsley}, {Leja}, {Charlot}, \& {Chevallard}}]{Whitler23}
{Whitler}, L., {Stark}, D.~P., {Endsley}, R., {et~al.} 2023, \mnras, 519, 5859

\bibitem[{{Woodrum} {et~al.}(2023){Woodrum}, {Rieke}, {Ji}, {Baker}, {Bhatawdekar}, {Bunker}, {Charlot}, {Curtis-Lake}, {Eisenstein}, {Hainline}, {Hausen}, {Helton}, {Hviding}, {Johnson}, {Robertson}, {Sun}, {Tacchella}, {Whitler}, {Williams}, \& {Willmer}}]{Woodrum23}
{Woodrum}, C., {Rieke}, M., {Ji}, Z., {et~al.} 2023, arXiv e-prints, arXiv:2310.18464

\bibitem[{{Wylezalek} {et~al.}(2022){Wylezalek}, {Vayner}, {Rupke}, {Zakamska}, {Veilleux}, {Ishikawa}, {Bertemes}, {Liu}, {Barrera-Ballesteros}, {Chen}, {Goulding}, {Greene}, {Hainline}, {Hamann}, {Heckman}, {Johnson}, {Lutz}, {L{\"u}tzgendorf}, {Mainieri}, {Maiolino}, {Nesvadba}, {Ogle}, \& {Sturm}}]{Wylezalek22}
{Wylezalek}, D., {Vayner}, A., {Rupke}, D. S.~N., {et~al.} 2022, \apjl, 940, L7

\bibitem[{{Xiao} {et~al.}(2023){Xiao}, {Oesch}, {Elbaz}, {Bing}, {Nelson}, {Weibel}, {Naidu}, {Daddi}, {Bouwens}, {Matthee}, {Wuyts}, {Chisholm}, {Brammer}, {Dickinson}, {Magnelli}, {Leroy}, {van Dokkum}, {Schaerer}, {Herard-Demanche}, {Barrufet}, {Endsley}, {Fudamoto}, {G{\'o}mez-Guijarro}, {Gottumukkala}, {Illingworth}, {Labbe}, {Magee}, {Marchesini}, {Maseda}, {Qin}, {Reddy}, {Shapley}, {Shivaei}, {Shuntov}, {Stefanon}, {Whitaker}, \& {Wyithe}}]{Xiao23}
{Xiao}, M., {Oesch}, P., {Elbaz}, D., {et~al.} 2023, arXiv e-prints, arXiv:2309.02492

\bibitem[{{Zibetti} {et~al.}(2009){Zibetti}, {Charlot}, \& {Rix}}]{Zibetti09}
{Zibetti}, S., {Charlot}, S., \& {Rix}, H.-W. 2009, \mnras, 400, 1181

\end{thebibliography}

\appendix \label{sec:appendix}

\onecolumn

\section{Star Formation Histories} \label{sec:app_sfh}
In \S\ref{sec:bagpipes}, we described the setup used within \textsc{Bagpipes} to model the SEDs. We use different SFH parameterisations. Here we specify the SFH forms and priors imposed, which can be found on Table \ref{tab:sfh_priors}. On top of the above-mentioned basis of four free parameters, we list here the additional ones.

\begin{table*}[h]
\centering
\caption{Different parametric forms of the star formation history that we use with \textsc{Bagpipes}. Each model has its own parameters and we specify the uniform priors used in this work.}
\begin{tabular}{l c}
\toprule
\toprule
SFH model & Priors  \\
\midrule
Constant & Maximum age $\in$ [1~Myr, 1~Gyr] \\ 
Log-normal & Age of the Universe at peak SF $\in$ [1~Myr, 1~Gyr] \\
 & Full width at half maximum SF $\in$ [0, 1~Gyr] \\
Exponentially Declining & Time since SFH began $\in$ [1~Myr, 1~Gyr] \\
& Timescale of decrease $\tau \in$ [0, 10~Gyr] \\
Double-Power Law & Falling slope index $\in$ [0, 10] \\
 & Rising slope index $\in$ [0, 10] \\
 & Age of the Universe at turnover $\tau \in$ [1~Myr, 1~Gyr] \\
\bottomrule
\end{tabular}
\label{tab:sfh_priors}
\end{table*}

\section{S/N Maps} \label{sec:app_sn}

\begin{figure}[h]
\centering
\includegraphics[width=\textwidth]{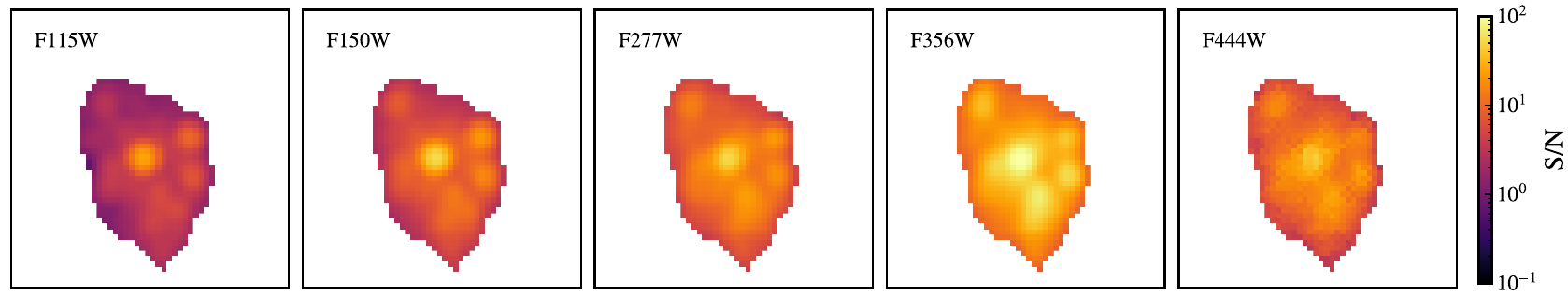}
\caption{Resulting S/N in all available NIRCam bands after imposing the S/N criteria discussed in \S\ref{sec:segmentation}.
\label{fig:app_sn}}
\end{figure}

In \S\ref{sec:segmentation}, we explained how the pixels are selected according to a S/N threshold in all bands. Figure~\ref{fig:app_sn} displays the resulting S/N maps per NIRCam band after the S/N criteria is applied.

\section{SED Fits} \label{sec:app_fits}

\subsection{Spatially-Integrated}

\begin{figure}[h]
\centering
\includegraphics[width=0.85\textwidth]{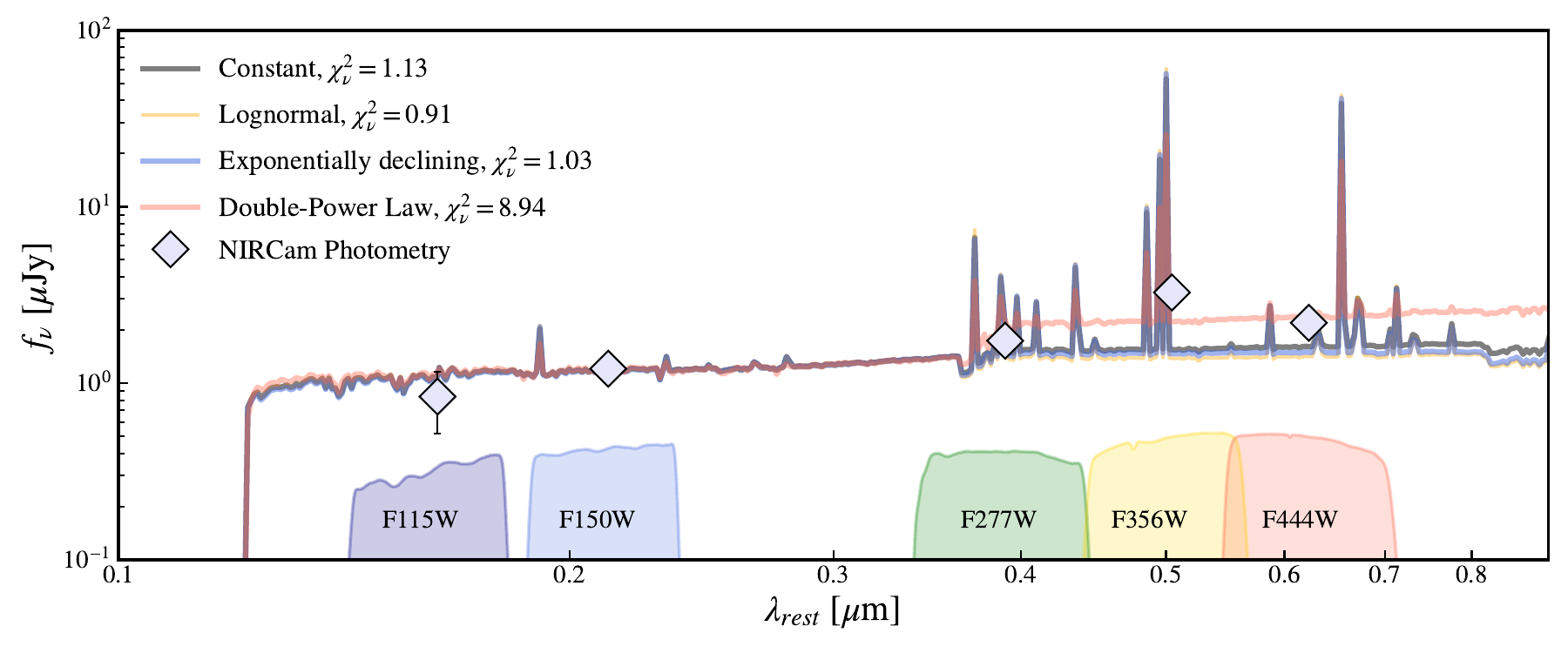}
\caption{Best fit models (solid coloured curves) for the various SFH parameterisations tested in the spatially-integrated photometry (diamonds). The NIRCam filter curves are displayed in the lower part.
\label{fig:app_fits}}
\end{figure}

\begin{figure}[h]
\centering
\includegraphics[width=0.85\textwidth]{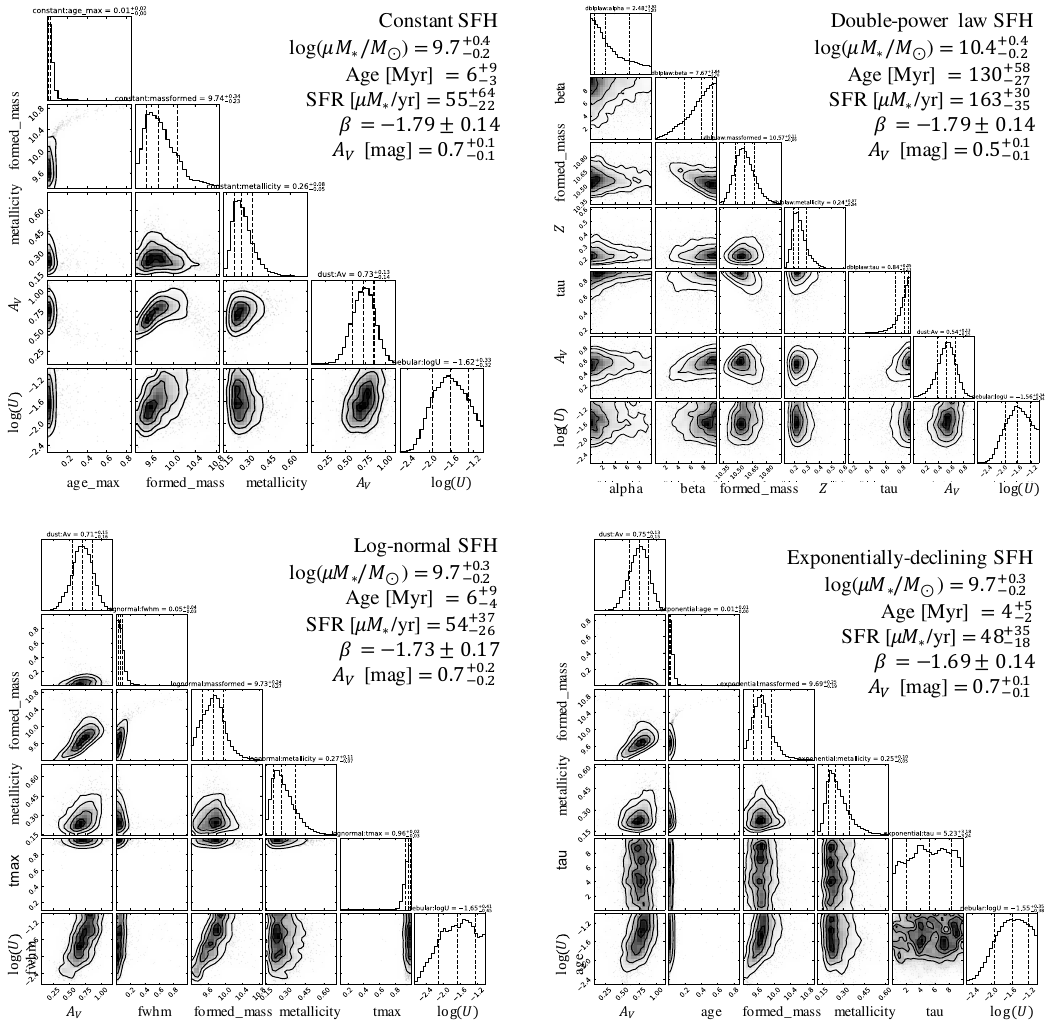}
\caption{Corner plots of the spatially-integrated fits using different SFH parameterisations: constant (top left), double-power law (top right), lognormal (bottom left), and exponentially declining (bottom right).
\label{fig:app_corner}}
\end{figure}

Here we present the different fits for the spatially-integrated photometry using the SFH shapes specified in the previous section. Figure~\ref{fig:app_fits} shows the best fit SED models for each SFH on top of the integrated NIRCam photometry, as well as the reduced chi squared values for each fit. Figure~\ref{fig:app_corner} displays the corner plots of all fits, as well as the inferred physical properties. As discussed in the caveats \S\ref{sec:caveats}, the SFHs with most free parameters such as the double-power law, show correlations and degeneracies between some of the fitted parameters. In terms of the SED fit, all models seem to fit the data, with the main difference being around the F356W and F444W bands. The DPL shows large continuum emission by old stellar populations, whereas the rest of models fit the photometric excess with strong \oiii\ and \nii\ emission lines instead, as discussed in \S\ref{sec:mass}.

\subsection{Spatially-Resolved}

\begin{figure}[t]
\centering
\includegraphics[width=0.9\textwidth]{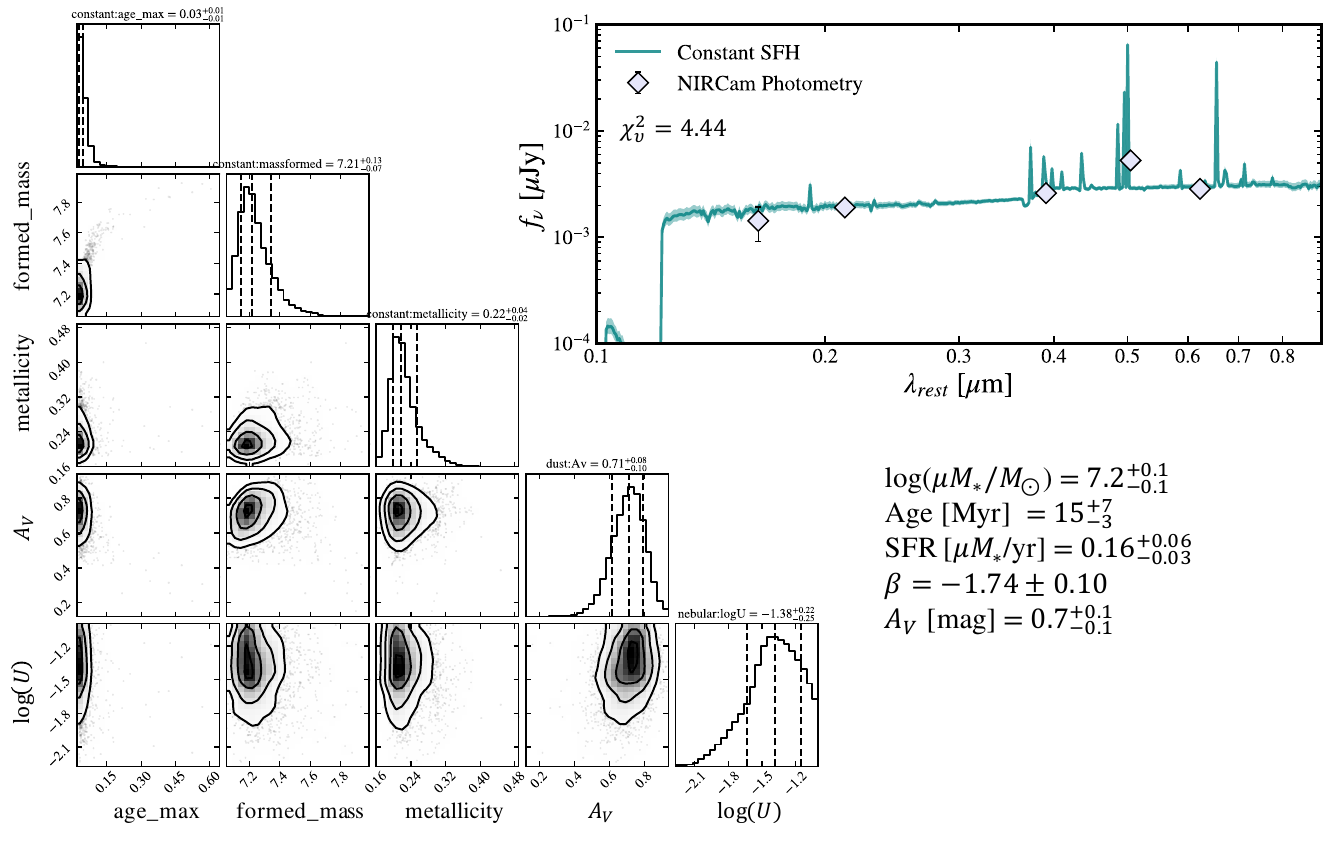}
\includegraphics[width=0.9\textwidth]{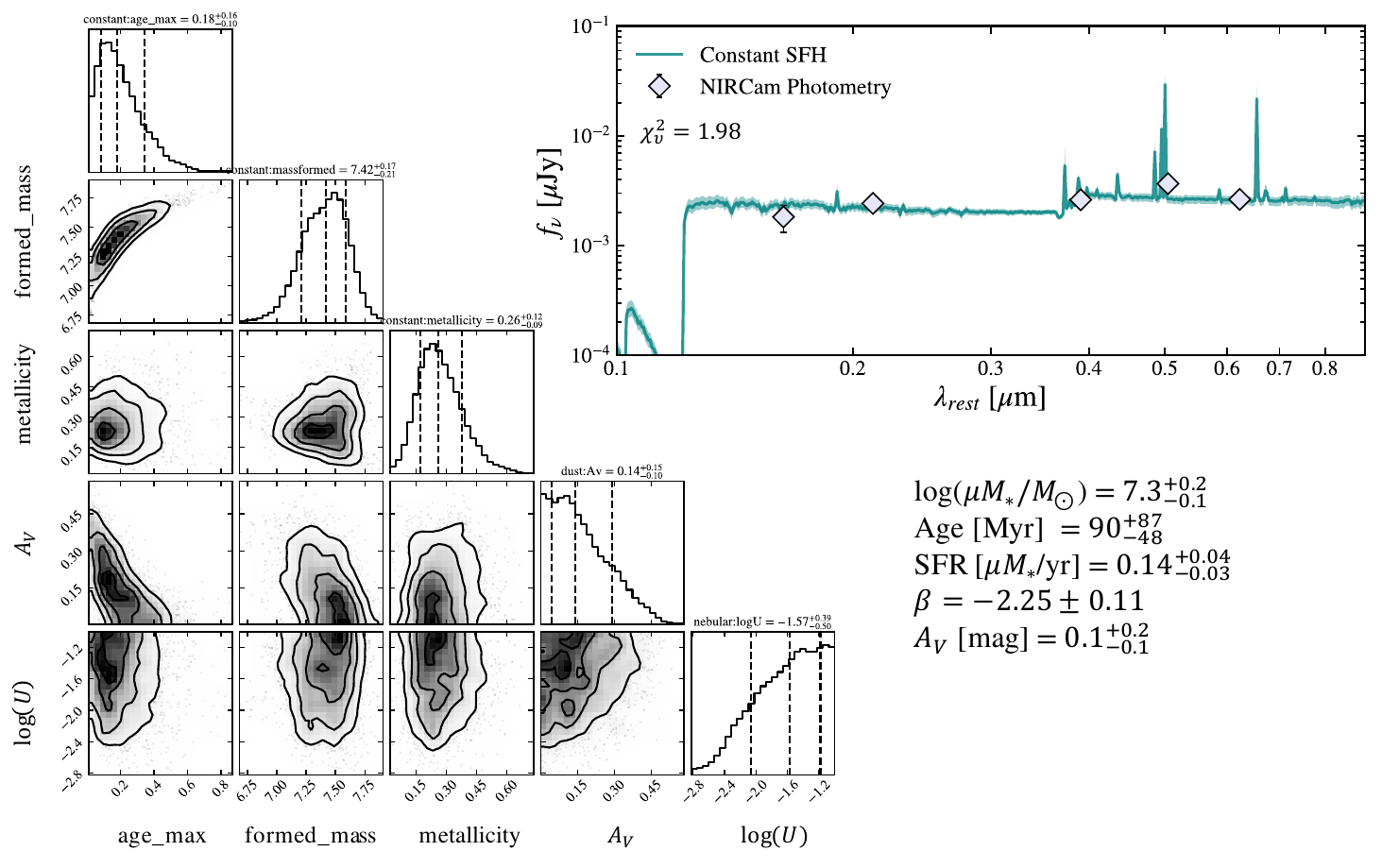}
\caption{Best fit models and corner plots for two representative pixels in the spatially-resolved analysis, a young (top) and an older (bottom) pixel. The NIRCam photometry for each pixel is indicated by the diamond symbols.
\label{fig:app_resfits}}
\end{figure}

Here we provide the fits and corner plots for two example pixels from the resolved maps presented in Figure~\ref{fig:resolved_maps}. One pixel corresponds to the extended region of older stellar populations, and the other corresponds to one of the centrally-located multiple young clumps. Figure~\ref{fig:app_resfits} shows the best fit SEDs and corresponding corner plots. The top pixel corresponds to the central region, with a young age of the stellar population of $15^{+7}_{-3}$~Myr. We can see that the F356W excess is fitted with strong \oiii\ and \nii\ line emission. The bottom pixel corresponds to the extended region of older stellar populations, displaying an age of $90^{+87}_{-48}$~Myr, with stronger stellar continuum and weaker emission lines fitting the F356W excess instead. We see that, particularly for the young pixel, the parameters are well constrained and we do not observe strong correlations or degeneracies. For the older pixel, the distributions of age, $A_V$ and log($U$) are broader, but we can still constrain the stellar mass. Both SED models are visibly different, with the young pixel displaying a bluer slope and stronger emission lines, and the old pixel having much redder UV slope and weak emission lines.

\end{document}